\documentclass[12pt,twoside]{article}

\usepackage{graphicx, epsfig}
\usepackage{amsfonts, amssymb, amsmath, amsthm}
\usepackage{latexsym}
\usepackage{cite}
\newcommand{\partdif}[2]{\ensuremath{ \frac{\partial #1}{\partial #2}}}

\begin{document}

\title{Extending a Hybrid Godunov Method for Radiation Hydrodynamics to Multiple Dimensions}
\author{Michael D Sekora \\ 
\textit{{\small Program in Applied and Computational Mathematics$^{\dagger}$}} \\ 
\textit{{\small Princeton University, Princeton, NJ 08544, USA}}}
\date{1 September 2010}
\maketitle

\begin{abstract}
\noindent
This paper presents a hybrid Godunov method for three-dimensional radiation hydrodynamics. The multidimensional technique outlined in this paper is an extension of the one-dimensional method that was developed by Sekora \& Stone 2009, 2010. The earlier one-dimensional technique was shown to preserve certain asymptotic limits and be uniformly well behaved from the photon free streaming (hyperbolic) limit through the weak equilibrium diffusion (parabolic) limit and to the strong equilibrium diffusion (hyperbolic) limit. This paper gives the algorithmic details for constructing a multidimensional method. A future paper will present numerical tests that demonstrate the robustness of the computational technique across a wide-range of parameter space.
\end{abstract}

\pagestyle{myheadings}
\markboth{M D Sekora}{Extending a Hybrid Godunov Method to Multiple Dimensions}


\section{Radiation Hydrodynamics}
\noindent
The purpose of this paper is to extend the ideas of Sekora \& Stone 2010 to radiation hydrodynamical problems in multiple dimensions. Therefore, this paper is a continuation of that earlier work, where the system of equations for radiation hydrodynamics were non-dimensionalized with respect to the material flow scale \cite{lowrie1999, lowrie2001}. This scaling gives two important parameters: $\mathbb{C} = c / a_{\infty}$ which measures relativistic effects and $\mathbb{P} = a_r T^4_{\infty} / \rho_{\infty}a^2_{\infty} $ which measures how radiation affects material dynamics. Additionally, $a_r = 8 \pi^5 k^4 / 15 c^3 h^3 = 7.57 \times 10^{-15} ~ erg ~ cm^{-3} ~ K^{-4}$ is a radiation constant, $T_{\infty}$ is a reference material temperature in the absence of radiation, and $\rho_{\infty}$ is a reference material density in the absence of radiation. The full system of equations for radiation hydrodynamics in the mixed frame that is correct to $\mathcal{O}(1/\mathbb{C})$ is:
\begin{equation}
\partdif{\rho}{t} + \nabla \cdot \left( \mathbf{m} \right) = 0 \label{eq:rh1} ,
\end{equation}
\begin{equation}
\partdif{\mathbf{m}}{t} + \nabla \cdot \left( \frac{\mathbf{m} \otimes \mathbf{m}}{\rho} \right) + \nabla p = -\mathbb{P} \left [ -\sigma_t \left( \mathbf{F_r} - \frac{ \mathbf{u}E_r + \mathbf{u} \cdot \mathsf{P_r} }{\mathbb{C}} \right) + \sigma_a \frac{\mathbf{u}}{\mathbb{C}} (T^4 - E_r) \right ] \label{eq:rh2} , 
\end{equation}
\begin{equation}
\partdif{E}{t} + \nabla \cdot \left( (E+p) \frac{\mathbf{m}}{\rho} \right) = -\mathbb{P} \mathbb{C} \left [ \sigma_a(T^4 - E_r) + (\sigma_a - \sigma_s) \frac{\mathbf{u}}{\mathbb{C}} \cdot \left( \mathbf{F_r} - \frac{ \mathbf{u}E_r + \mathbf{u} \cdot \mathsf{P_r} }{\mathbb{C}} \right) \right ] \label{eq:rh3} , 
\end{equation}
\begin{equation}
\partdif{E_r}{t} + \mathbb{C} \nabla \cdot \mathbf{F_r} = \mathbb{C} \left [ \sigma_a(T^4 - E_r) + (\sigma_a - \sigma_s) \frac{\mathbf{u}}{\mathbb{C}} \cdot \left( \mathbf{F_r} - \frac{ \mathbf{u}E_r + \mathbf{u} \cdot \mathsf{P_r} }{\mathbb{C}} \right) \right ] \label{eq:rh4} , 
\end{equation}
\begin{equation}
\partdif{ \mathbf{F_r} }{t} + \mathbb{C} \nabla \cdot \mathsf{P_r} = \mathbb{C} \left [ -\sigma_t \left( \mathbf{F_r} - \frac{ \mathbf{u}E_r + \mathbf{u} \cdot \mathsf{P_r} }{\mathbb{C}} \right) + \sigma_a \frac{\mathbf{u}}{\mathbb{C}} (T^4 - E_r) \right ] \label{eq:rh5} ,
\end{equation}
\begin{equation}
\mathsf{P_r} = \mathsf{f} E_r ~~ \textrm{(closure relation)} \label{eq:rad_closure} , ~~ \mathsf{f} = \frac{1 - \chi}{2} I + \frac{3 \chi - 1}{2} \mathbf{n} \otimes \mathbf{n} .
\end{equation}

\noindent
For the material quantities, $\rho$ is density, $\mathbf{m}$ is momentum density, $p = (\gamma-1)e $ is pressure, $E$ is energy density, and $T$ is temperature. For the radiation quantities, $E_r$ is energy density, $\mathbf{F_r}$ is flux, $\mathsf{P_r}$ is pressure, $\mathsf{f}$ is the variable tensor Eddington factor that is used to close the hierarchy of radiation transport moment equations, $\chi$ is a parameter, and $\mathbf{n}$ is a unit normal vector aligned with the radiative flux \cite{levermore1984, heracles2007}. In the free streaming limit (optically thin regime), $\chi \rightarrow 1$ such that $\mathsf{f} \rightarrow \mathbf{n} \otimes \mathbf{n}$. Yet, in the equilibrium diffusion limit (optically thick regime), $\chi \rightarrow 1/3$ such that $\mathsf{f} \rightarrow \frac{1}{3} I$, where $I$ is the identity matrix. In the above modified Mihalas-Klein source terms, $\sigma_a$ is the absorption cross section, $\sigma_s$ is the scattering cross section, and $\sigma_t = \sigma_a + \sigma_s$ is the total cross section \cite{lowrie1999, lowrie2001, mk1982}. \\

\noindent
The numerical approach of the hybrid Godunov method rests on understanding the balance law form of the above system of equations, where:
\begin{equation}
\partdif{U}{t} + \partdif{F(U)}{x} + \partdif{G(U)}{y} + \partdif{H(U)}{z} = S(U) \label{eq:cons_law} ,
\end{equation}
\begin{equation}
U = 
\left( \begin{array}{c} \rho    \\
                        m_{x}   \\
                        m_{y}   \\
                        m_{z}   \\
                        E       \\
                        E_{r}   \\
                        F_{r,x} \\
                        F_{r,y} \\
                        F_{r,z} \end{array} \right) , ~~~~
F(U) = 
\left( \begin{array}{c} m_{x} \\
                        \frac{m_{x}^2}{\rho} + p  \\
                        \frac{m_{x} m_{y}}{\rho}  \\
                        \frac{m_{x} m_{z}}{\rho}  \\
                        (E + p)\frac{m_{x}}{\rho} \\
                        \mathbb{C} F_{r,x} \\
                        \mathbb{C} f_{xx} E_r \\
                        \mathbb{C} f_{yx} E_r \\
                        \mathbb{C} f_{zx} E_r \end{array} \right) , ~~~~
G(U) = 
\left( \begin{array}{c} m_{y} \\
                        \frac{m_{y} m_{x}}{\rho}  \\
                        \frac{m_{y}^2}{\rho} + p  \\
                        \frac{m_{y} m_{z}}{\rho}  \\
                        (E + p)\frac{m_{y}}{\rho} \\
                        \mathbb{C} F_{r,y} \\
                        \mathbb{C} f_{xy} E_r \\
                        \mathbb{C} f_{yy} E_r \\
                        \mathbb{C} f_{zy} E_r \end{array} \right) , 
\end{equation}
\begin{equation}
H(U) = 
\left( \begin{array}{c} m_{z} \\
                        \frac{m_{z} m_{x}}{\rho}  \\
                        \frac{m_{z} m_{y}}{\rho}  \\
                        \frac{m_{z}^2}{\rho} + p  \\
                        (E + p)\frac{m_{z}}{\rho} \\
                        \mathbb{C} F_{r,z} \\
                        \mathbb{C} f_{xz} E_r \\
                        \mathbb{C} f_{yz} E_r \\
                        \mathbb{C} f_{zz} E_r \end{array} \right) , ~~~~
S(U) = 
\left( \begin{array}{c} 0 \\
                        -\mathbb{P} S^{F_x} \\
                        -\mathbb{P} S^{F_y} \\
                        -\mathbb{P} S^{F_z} \\
                        -\mathbb{P} \mathbb{C} S^E \\
                        \mathbb{C} S^E \\
                        \mathbb{C} S^{F_x} \\ 
                        \mathbb{C} S^{F_y} \\
                        \mathbb{C} S^{F_z} \end{array} \right) , 
\end{equation}

\noindent
where:
\begin{eqnarray}
S^{F_x} &=& -\sigma_t \left( F_{r,x} - \frac{E_r}{\rho \mathbb{C}} ( m_{x} + m_{x} f_{xx} + m_{y} f_{yx} + m_{z} f_{zx} ) \right) + \sigma_a \frac{m_{x}}{\rho \mathbb{C}} (T^4 - E_r) , \\ 
S^{F_y} &=& -\sigma_t \left( F_{r,y} - \frac{E_r}{\rho \mathbb{C}} ( m_{y} + m_{x} f_{xy} + m_{y} f_{yy} + m_{z} f_{zy} ) \right) + \sigma_a \frac{m_{y}}{\rho \mathbb{C}} (T^4 - E_r) , \\ 
S^{F_z} &=& -\sigma_t \left( F_{r,z} - \frac{E_r}{\rho \mathbb{C}} ( m_{z} + m_{x} f_{xz} + m_{y} f_{yz} + m_{z} f_{zz} ) \right) + \sigma_a \frac{m_{z}}{\rho \mathbb{C}} (T^4 - E_r) , \\ 
S^{E}   &=& \sigma_a(T^4 - E_r) + (\sigma_a - \sigma_s) \frac{m_{x}}{\rho \mathbb{C}} \left( F_{r,x} - \frac{E_r}{\rho \mathbb{C}} ( m_{x} + m_{x} f_{xx} + m_{y} f_{yx} + m_{z} f_{zx} ) \right) \\
~       &~& ~~~~~~~~~~~~~~~~    + (\sigma_a - \sigma_s) \frac{m_{y}}{\rho \mathbb{C}} \left( F_{r,y} - \frac{E_r}{\rho \mathbb{C}} ( m_{y} + m_{x} f_{xy} + m_{y} f_{yy} + m_{z} f_{zy} ) \right) \nonumber \\ 
~       &~& ~~~~~~~~~~~~~~~~    + (\sigma_a - \sigma_s) \frac{m_{z}}{\rho \mathbb{C}} \left( F_{r,z} - \frac{E_r}{\rho \mathbb{C}} ( m_{z} + m_{x} f_{xz} + m_{y} f_{yz} + m_{z} f_{zz} ) \right) . \nonumber
\end{eqnarray}

\noindent
The quasi-linear form of this system of hyperbolic balance laws is:
\begin{equation}
\partdif{U}{t} + A_{x} \partdif{U}{x} + A_{y} \partdif{U}{y} + A_{z} \partdif{U}{z} = S(U) \label{eq:cons_law_quasi} ,
\end{equation}
{\footnotesize{ \begin{equation}
A_{x} = \left( \begin{array}{ccccccccc} 
0                                                 & 1                       & 0             & 0             & 0          & 0 & 0 & 0 & 0 \\
\frac{\gamma-1}{2} V^2 - u^2                      & -(\gamma-3) u           & -(\gamma-1) v & -(\gamma-1) w & (\gamma-1) & 0 & 0 & 0 & 0 \\
-u v                                              & v                       & u             & 0             & 0          & 0 & 0 & 0 & 0 \\
-u w                                              & w                       & 0             & u             & 0          & 0 & 0 & 0 & 0 \\
u\left( \frac{\gamma-1}{2} V^2-\tilde{H} \right)  & \tilde{H}-(\gamma-1)u^2 & -(\gamma-1)uv & -(\gamma-1)uw & \gamma u   & 0 & 0 & 0 & 0 \\
0                                                 & 0                       & 0             & 0             & 0          & 0 & \mathbb{C} & 0 & 0 \\
0                                                 & 0                       & 0             & 0             & 0          & \mathbb{C} f_{xx} & 0 & 0 & 0 \\
0                                                 & 0                       & 0             & 0             & 0          & \mathbb{C} f_{yx} & 0 & 0 & 0 \\
0                                                 & 0                       & 0             & 0             & 0          & \mathbb{C} f_{zx} & 0 & 0 & 0 \end{array} \right) \label{eq:cons_law_jacobian_ax} 
\end{equation}

\begin{equation}
A_{y} = \left( \begin{array}{ccccccccc} 
0                                                 & 0                       & 1             & 0             & 0          & 0 & 0 & 0 & 0 \\
-u v                                              & v                       & u             & 0             & 0          & 0 & 0 & 0 & 0 \\
\frac{\gamma-1}{2} V^2 - v^2                      & -(\gamma-1) u           & -(\gamma-3) v & -(\gamma-1) w & (\gamma-1) & 0 & 0 & 0 & 0 \\
-v w                                              & 0                       & w             & v             & 0          & 0 & 0 & 0 & 0 \\
v\left( \frac{\gamma-1}{2} V^2-\tilde{H} \right)  & -(\gamma-1)uv & \tilde{H}-(\gamma-1)v^2 & -(\gamma-1)vw & \gamma v   & 0 & 0 & 0 & 0 \\
0                                                 & 0                       & 0             & 0             & 0          & 0 & 0 & \mathbb{C} & 0 \\
0                                                 & 0                       & 0             & 0             & 0          & \mathbb{C} f_{xy} & 0 & 0 & 0 \\
0                                                 & 0                       & 0             & 0             & 0          & \mathbb{C} f_{yy} & 0 & 0 & 0 \\
0                                                 & 0                       & 0             & 0             & 0          & \mathbb{C} f_{zy} & 0 & 0 & 0 \end{array} \right) \label{eq:cons_law_jacobian_ay} 
\end{equation}

\begin{equation}
A_{z} = \left( \begin{array}{ccccccccc} 
0                                                 & 0                       & 0             & 1             & 0          & 0 & 0 & 0 & 0 \\
-u w                                              & w                       & 0             & u             & 0          & 0 & 0 & 0 & 0 \\
-v w                                              & 0                       & w             & v             & 0          & 0 & 0 & 0 & 0 \\
\frac{\gamma-1}{2} V^2 - w^2                      & -(\gamma-1) u           & -(\gamma-1) v & -(\gamma-3) w & (\gamma-1) & 0 & 0 & 0 & 0 \\
w\left( \frac{\gamma-1}{2} V^2-\tilde{H} \right)  & -(\gamma-1)uw & -(\gamma-1)vw & \tilde{H}-(\gamma-1)w^2 & \gamma w   & 0 & 0 & 0 & 0 \\
0                                                 & 0                       & 0             & 0             & 0          & 0 & 0 & 0 & \mathbb{C} \\
0                                                 & 0                       & 0             & 0             & 0          & \mathbb{C} f_{xz} & 0 & 0 & 0 \\
0                                                 & 0                       & 0             & 0             & 0          & \mathbb{C} f_{yz} & 0 & 0 & 0 \\
0                                                 & 0                       & 0             & 0             & 0          & \mathbb{C} f_{zz} & 0 & 0 & 0 \end{array} \right) \label{eq:cons_law_jacobian_az} 
\end{equation} }}

\noindent
Here, $u = m_x/\rho$, $v = m_y/\rho$, and $w = m_z/\rho$ are the velocities in the $x$, $y$, and $z$ directions, respectively. $V^2 = u^2 + v^2 + w^2$ and $\tilde{H} = \frac{\gamma E}{\rho} - \frac{\gamma-1}{2}V^2$ is specific enthalpy. The Jacobians $A_x$, $A_y$, and $A_z$ have eigenvalues: $\lambda_x = \{ 0, u, u \pm a, \pm f_{xx}^{1/2} \mathbb{C} \}$, $\lambda_y = \{ 0, v, v \pm a, \pm f_{yy}^{1/2} \mathbb{C} \}$, and $\lambda_z = \{ 0, w, w \pm a, \pm f_{zz}^{1/2} \mathbb{C} \}$, respectively. However, one must account for how the stiff momentum and energy source terms affect the hyperbolic structure.


\section{Overview of the Multidimensional Algorithm}
\noindent
In radiation hydrodynamics, there are three important dynamical scales: the speed of sound (material flow), speed of light (radiation flow), and speed at which the source terms interact. Given such variation, one desires a numerical technique that treats the material flow explicitly, radiation flow implicitly, and source terms semi-implicitly.

\subsection{Effective CFL Condition}
\noindent
Our primary interest when solving radiation hydrodynamical problems is higher-order resolution of material quantities while advancing the overall algorithm according to an effective CFL condition. This temporal advancement is associated with an effective CTU (corner transport upwind) scheme \cite{colella1990}:
\begin{equation}
\Delta t = \frac{\nu}{ \max_{i,j,k} \left \{ \frac{ |u(i,j,k)| + a_{\textrm{eff}}(i,j,k) }{\Delta x} , \frac{ |v(i,j,k)| + a_{\textrm{eff}}(i,j,k) }{\Delta y} , \frac{ |w(i,j,k)| + a_{\textrm{eff}}(i,j,k) }{\Delta z} \right \} } ,
\end{equation}

\noindent 
where $\Delta t$ is the time step and $\nu \in [0,1]$ is the CFL number. $\Delta x = (x_{\max}-x_{\min}) / N_{cell}^{x}$, $\Delta y = (y_{\max}-y_{\min}) / N_{cell}^{y}$, and $\Delta z = (z_{\max}-z_{\min}) / N_{cell}^{z}$ are the spatial resolutions for a given number of computational grid cells in the $x$, $y$, and $z$ directions, respectively. $\max_{i,j,k} \left \{ |u(i,j,k)| + a_{\textrm{eff}}(i,j,k) , |v(i,j,k)| + a_{\textrm{eff}}(i,j,k) , |w(i,j,k)| + a_{\textrm{eff}}(i,j,k) \right \}$ is the maximum material wave speed over all grid cells. Furthermore, $a_{\textrm{eff}}$ is an estimate of the radiation modified sound speed which is obtained by carrying out an effective eigen-analysis of the material Jacobian. This analysis is presented in a later section. One chooses a CTU-type time step over other alternatives (e.g., donor-cell time step) because of how one couples transport in the corners of the computational grid cells. For the duration of this paper, one assumes that the grid cells are cubic $(\Delta x = \Delta y = \Delta z)$.

\subsection{Algorithmic Steps}
\noindent
After defining $\Delta t$, the algorithm loops over the following steps:
\begin{enumerate}
\item Backward Euler Upwinding Scheme - implicitly advances the radiation quantities from time $t_{n}$ to time $t_{n+1}$: 
\begin{equation}
(E_r^{n}, F_{r,x}^{n}, F_{r,y}^{n}, F_{r,z}^{n}) \rightarrow (E_r^{n+1},F_{r,x}^{n+1}, F_{r,y}^{n+1}, F_{r,z}^{n+1}) \nonumber
\end{equation}
\item Modified Godunov Predictor Scheme - couples stiff source term effects to the hyperbolic structure of the balance laws for the material quantities and uses effective piecewise linear extrapolation to spatially reconstruct material quantities at the left/right sides of cell interfaces in the $x$, $y$, and $z$ directions $\left \{i \pm 1/2, j \pm 1/2, k \pm 1/2 \right \}$: 
\begin{equation}
\tilde{U}_{L/R,i+1/2,j,k}^{m,n+1/2}, ~~~~ \tilde{U}_{L/R,i,j+1/2,k}^{m,n+1/2}, ~~~~ \tilde{U}_{L/R,i,j,k+1/2}^{m,n+1/2} \nonumber
\end{equation}
\item Flux Function - evaluates the passage of material across cell interfaces using the left/right material states as well as an approximate Riemann solver to obtain: 
\begin{eqnarray}
\tilde{F}_{i+1/2,j,k}^{m} &=& F(\mathcal{R}(\tilde{U}_{L,i+1/2,j,k}^{m},\tilde{U}_{R,i+1/2,j,k}^{m})) \nonumber \\
\tilde{G}_{i,j+1/2,k}^{m} &=& G(\mathcal{R}(\tilde{U}_{L,i,j+1/2,k}^{m},\tilde{U}_{R,i,j+1/2,k}^{m})) \nonumber \\
\tilde{H}_{i,j,k+1/2}^{m} &=& H(\mathcal{R}(\tilde{U}_{L,i,j,k+1/2}^{m},\tilde{U}_{R,i,j,k+1/2}^{m})) \nonumber
\end{eqnarray}
\item CTU Correction - accounts for how the material quantities at the left/right sides of cell interfaces in one spatial direction are affected by the fluxes in the other two spatial directions: 
\begin{equation}
\tilde{U}_{L/R}^{m,n+1/2} \rightarrow U_{L/R}^{m,n+1/2} \nonumber
\end{equation}
\item Flux Function - evaluates the passage of material across cell interfaces using the corrected left/right material states and the algorithmic machinery of Step 3 above
\item Modified Godunov Corrector Scheme - semi-implicitly advances the material quantities from time $t_{n}$ to time $t_{n+1}$: 
\begin{equation}
(\rho^{n},m_{x}^{n},m_{y}^{n},m_{z}^{n},E^{n}) \rightarrow (\rho^{n+1},m_{x}^{n+1},m_{y}^{n+1},m_{z}^{n+1},E^{n+1}) \nonumber
\end{equation}
\item Apply boundary conditions
\item Compute next time step $\Delta t$
\end{enumerate}

\noindent
In the above expressions, $U$, $U^{r}$, and $U^{m}$ represent all of the conserved quantities, radiation quantities $(E_{r}$,$F_{r,x}$,$F_{r,y}$,$F_{r,z})$, and material quantities $(\rho$,$m_x$,$m_y$, $m_z$,$E)$, respectively. $F_{i+1/2,j,k}$, $G_{i,j+1/2,k}$, and $H_{i,j,k+1/2}$ are fluxes directed across cell faces in the $x$, $y$, and $z$ directions, where $\mathcal{R}$ represents the solution of a Riemann problem. The tilde that is above some of the left/right material states and fluxes designates quantities that have not yet been adjusted by the CTU correction. Cell centers are defined by three indices $(i,j,k)$, such that $i \pm 1/2$, $j \pm 1/2$, and $k \pm 1/2$ represent the location of a cell interface to the right/left of $i$, $j$, and $k$, respectively. $n$ designates the time discretization. Details about each step are explained in later sections. Lastly, the one-dimensional hybrid Godunov method of \cite{sekora2010} was shown to be consistent, stable, and accurate as well as coarsely gridded, well-balanced, and having some asymptotic preserving properties. For reasons cited in \cite{sekora2010}, these numerical properties should be able to be extended to the multidimensional algorithm. However, future tests will justify these claims.


\section{Backward Euler Upwinding Scheme}
\noindent
This section presents the implicit scheme that advances the radiation quantities $U^{r}$ according to the material flow scale. The stability of explicit schemes (e.g., Godunov-type methods) is governed by the CFL condition which restricts the allowable time step according to the fastest characteristic speed. However, if a hyperbolic system consists of multiscale waves (e.g., radiation hydrodynamics where $c / a_{\infty} \sim 10^{6}$), then explicit schemes can become inefficient. For these types of problems, implicit methods are useful. Following \cite{duraisamy2007, gottlieb2001}, one can construct implicit flux functions to approximate integrals at cell interfaces. In this context, a HLLE framework was implemented.

\subsection{HLLE Framework}
\noindent
The HLLE scheme is based on estimating the minimum and maximum wave speeds $(s_{\min}, s_{\max})$. These quantities are uniquely defined for each Riemann problem that is associated with each interface of a computational grid cell. The numerical flux in one direction is calculated using the following formula:
\begin{equation}
F^{\textrm{HLLE}} (\mathcal{R}(U_{L},U_{R})) = \frac{1}{2} \left( \left( 1 + C^{s} \right) \left( F(U_{L}) - s_{\min} U_{L} \right) + \left( 1 - C^{s} \right) \left( F(U_{R}) - s_{\max} U_{R} \right) \right) . \label{eq:HLLE1} 
\end{equation}

\noindent 
where $C^{s} = (s_{\max} + s_{\min}) / (s_{\max} - s_{\min})$. Defining the left/right states of the Riemann problem according to a first-order accurate (piecewise constant) reconstruction, forces the HLLE flux function to take the following form in each of the spatial directions:
\begin{eqnarray}
F_{i+1/2,j,k} (\mathcal{R}(U_{L,i+1/2,j,k},U_{R,i+1/2,j,k})) & \rightarrow & F_{i+1/2,j,k}^{\textrm{HLLE}} (\mathcal{R}(U_{i,j,k},U_{i+1,j,k})) , \\
G_{i,j+1/2,k} (\mathcal{R}(U_{L,i,j+1/2,k},U_{R,i,j+1/2,k})) & \rightarrow & G_{i,j+1/2,k}^{\textrm{HLLE}} (\mathcal{R}(U_{i,j,k},U_{i,j+1,k})) , \\
H_{i,j,k+1/2} (\mathcal{R}(U_{L,i,j,k+1/2},U_{R,i,j,k+1/2})) & \rightarrow & H_{i,j,k+1/2}^{\textrm{HLLE}} (\mathcal{R}(U_{i,j,k},U_{i,j,k+1})) ,
\end{eqnarray}

\noindent
A first-order accurate, backward Euler-type algorithm was used because of total variation diminishing (TVD) conditions. These issues were explored by \cite{duraisamy2007, sekora2010}. One makes the above explicit HLLE scheme implicit by defining the variables in the flux and source terms to be at time $t_{n+1}$ such that the exact integral formulation of the conservative differencing is:
\begin{eqnarray}
U_{i,j,k}^{n+1} = U_{i,j,k}^{n} &-& \frac{\Delta t}{\Delta x} \left( F_{i+1/2,j,k}^{\textrm{HLLE}}(\mathcal{R}(U_{i,j,k}^{n+1},U_{i+1,j,k}^{n+1})) - F_{i-1/2,j,k}^{\textrm{HLLE}}(\mathcal{R}(U_{i-1,j,k}^{n+1},U_{i,j,k}^{n+1})) \right) \label{eq:con_diff_imp} \\
                                &-& \frac{\Delta t}{\Delta y} \left( G_{i,j+1/2,k}^{\textrm{HLLE}}(\mathcal{R}(U_{i,j,k}^{n+1},U_{i,j+1,k}^{n+1})) - G_{i,j-1/2,k}^{\textrm{HLLE}}(\mathcal{R}(U_{i,j-1,k}^{n+1},U_{i,j,k}^{n+1})) \right) \nonumber \\
                                &-& \frac{\Delta t}{\Delta z} \left( H_{i,j,k+1/2}^{\textrm{HLLE}}(\mathcal{R}(U_{i,j,k}^{n+1},U_{i,j,k+1}^{n+1})) - H_{i,j,k-1/2}^{\textrm{HLLE}}(\mathcal{R}(U_{i,j,k-1}^{n+1},U_{i,j,k}^{n+1})) \right) \nonumber \\
                                &+& \Delta t S(U_{i,j,k}^{n+1})  \nonumber ,
\end{eqnarray}
\begin{eqnarray}
F_{i+1/2,j,k}^{\textrm{HLLE}} = &~& \frac{1}{2} \left( 1 + C_{i+1/2,j,k}^{s,x} \right) \left( F(U_{i,j,k}^{n+1}) - s_{\min}^{x} U_{i,j,k}^{n+1} \right) \label{eq:hlle_imp_f} \\
                                &+& \frac{1}{2} \left( 1 - C_{i+1/2,j,k}^{s,x} \right) \left( F(U_{i+1,j,k}^{n+1}) - s_{\max}^{x} U_{i+1,j,k}^{n+1} \right) \nonumber , \\
G_{i,j+1/2,k}^{\textrm{HLLE}} = &~& \frac{1}{2} \left( 1 + C_{i,j+1/2,k}^{s,y} \right) \left( G(U_{i,j,k}^{n+1}) - s_{\min}^{y} U_{i,j,k}^{n+1} \right) \label{eq:hlle_imp_g} \\
                                &+& \frac{1}{2} \left( 1 - C_{i,j+1/2,k}^{s,y} \right) \left( G(U_{i,j+1,k}^{n+1}) - s_{\max}^{y} U_{i,j+1,k}^{n+1} \right) \nonumber  , \\
H_{i,j,k+1/2}^{\textrm{HLLE}} = &~& \frac{1}{2} \left( 1 + C_{i,j,k+1/2}^{s,z} \right) \left( H(U_{i,j,k}^{n+1}) - s_{\min}^{z} U_{i,j,k}^{n+1} \right) \label{eq:hlle_imp_h} \\
                                &+& \frac{1}{2} \left( 1 - C_{i,j,k+1/2}^{s,z} \right) \left( H(U_{i,j,k+1}^{n+1}) - s_{\max}^{z} U_{i,j,k+1}^{n+1} \right) \nonumber .
\end{eqnarray}

\subsection{Applying the Backward Euler HLLE Scheme}
\noindent
If one considers only the radiation part of the equations for radiation hydrodynamics (Equations \ref{eq:rh4} and \ref{eq:rh5}) \cite{sekora2009, sekora2010}, then the variables, fluxes, and source terms are:
\begin{equation}
U^{r} = 
\left( \begin{array}{c} E_{r}   \\
                        F_{r,x} \\
                        F_{r,y} \\
                        F_{r,z} \end{array} \right) , ~~~~
F^{r}(U) = 
\left( \begin{array}{c} \mathbb{C} F_{r,x} \\
                        \mathbb{C} f_{xx} E_r \\
                        \mathbb{C} f_{yx} E_r \\
                        \mathbb{C} f_{zx} E_r \end{array} \right) , ~~~~
G^{r}(U) = 
\left( \begin{array}{c} \mathbb{C} F_{r,y} \\
                        \mathbb{C} f_{xy} E_r \\
                        \mathbb{C} f_{yy} E_r \\
                        \mathbb{C} f_{zy} E_r \end{array} \right) , \label{eq:radsub1}
\end{equation}
\begin{equation}
H^{r}(U) = 
\left( \begin{array}{c} \mathbb{C} F_{r,z} \\
                        \mathbb{C} f_{xz} E_r \\
                        \mathbb{C} f_{yz} E_r \\
                        \mathbb{C} f_{zz} E_r \end{array} \right) , ~~~~
S^{r}(U) = 
\left( \begin{array}{c} \mathbb{C} S^E \\
                        \mathbb{C} S^{F_x} \\ 
                        \mathbb{C} S^{F_y} \\
                        \mathbb{C} S^{F_z} \end{array} \right) , \label{eq:radsub2}
\end{equation}

\noindent
where the eigenvalues of the radiation subsystem in the free streaming limit $( \sigma_a, \sigma_t \sim \mathcal{O}(\epsilon) )$ are $\lambda_x = \{ 0, \pm f_{xx}^{1/2} \mathbb{C} \}$, $\lambda_y = \{ 0,  \pm f_{yy}^{1/2} \mathbb{C} \}$, and $\lambda_z = \{ 0, \pm f_{zz}^{1/2} \mathbb{C} \}$ for each of the spatial directions. Given that the HLLE scheme uses minimum and maximum wave speeds to compute fluxes at cell interfaces, one defines the following equations:
\begin{equation}
s_{\min}^{x}        = \lambda_{x,L}^{-}(i  ,j,k) = - f_{xx}(i  ,j,k)^{1/2} \mathbb{C} , ~~~~
s_{\max}^{x}        = \lambda_{x,R}^{+}(i+1,j,k) =   f_{xx}(i+1,j,k)^{1/2} \mathbb{C} , \nonumber
\end{equation}
\begin{equation}
C_{i+1/2,j,k}^{s,x} = \frac{f_{xx}(i+1,j,k)^{1/2} - f_{xx}(i,j,k)^{1/2}}{f_{xx}(i+1,j,k)^{1/2} + f_{xx}(i,j,k)^{1/2}} , \nonumber
\end{equation}
\begin{equation}
s_{\min}^{y}        = \lambda_{y,L}^{-}(i,j  ,k) = - f_{yy}(i,j  ,k)^{1/2} \mathbb{C} , ~~~~
s_{\max}^{y}        = \lambda_{y,R}^{+}(i,j+1,k) =   f_{yy}(i,j+1,k)^{1/2} \mathbb{C} , \nonumber
\end{equation}
\begin{equation}
C_{i,j+1/2,k}^{s,y} = \frac{f_{yy}(i,j+1,k)^{1/2} - f_{yy}(i,j,k)^{1/2}}{f_{yy}(i,j+1,k)^{1/2} + f_{yy}(i,j,k)^{1/2}} , \nonumber
\end{equation}
\begin{equation}
s_{\min}^{z}        = \lambda_{z,L}^{-}(i,j,k  ) = - f_{zz}(i,j,k  )^{1/2} \mathbb{C} , ~~~~
s_{\max}^{z}        = \lambda_{z,R}^{+}(i,j,k+1) =   f_{zz}(i,j,k+1)^{1/2} \mathbb{C} , \nonumber
\end{equation}
\begin{equation}
C_{i,j,k+1/2}^{s,z} = \frac{f_{zz}(i,j,k+1)^{1/2} - f_{zz}(i,j,k)^{1/2}}{f_{zz}(i,j,k+1)^{1/2} + f_{zz}(i,j,k)^{1/2}} . \nonumber
\end{equation}

\noindent
Here, $f_{xx}$, $f_{yy}$, and $f_{zz}$ arise from the closure relation $\mathsf{P_{r}} = \mathsf{f} E_{r}$ and is either a user defined quantity or obtained by solving the radiation transport equation. If $\mathsf{f}$ varies spatially, then $C^{s,x}$, $C^{s,y}$, and $C^{s,z}$ are non-zero. Defining or computing $\mathsf{f}(\mathbf{x},t)$ precedes the backward Euler update of $U^{r}$. However, if $\mathsf{f}$ is assumed to be spatially and temporally constant, then $C^{s,x}$,$C^{s,y}$,$C^{s,z} = 0$. Future work will update $\mathsf{f}(\mathbf{x},t)$ at each iteration by solving the radiation transport equation.

\subsection{Matrix Equation for the Radiation Components}
\noindent
Inputting $U^{r,n+1}$, $F^{r}(U^{r,n+1})$, $G^{r}(U^{r,n+1})$, $H^{r}(U^{r,n+1})$ and $S^{r}(U^{m,n},U^{r,n+1})$ into Equations \ref{eq:con_diff_imp}-\ref{eq:hlle_imp_h} gives the following four implicit difference equations:

{\tiny {\begin{eqnarray}
& ~ & E_{r  }^{n+1}(i,j,k-1) \left[ -d_{1} \left( 1 + C_{i,j,k-1/2}^{s,z} \right) f_{zz}(i,j,k-1)^{1/2} \right] \\
& + & F_{r,z}^{n+1}(i,j,k-1) \left[ -d_{1} \left( 1 + C_{i,j,k-1/2}^{s,z} \right) \right] \nonumber \\
& + & E_{r  }^{n+1}(i,j-1,k) \left[ -d_{1} \left( 1 + C_{i,j-1/2,k}^{s,y} \right) f_{yy}(i,j-1,k)^{1/2} \right] \nonumber \\
& + & F_{r,y}^{n+1}(i,j-1,k) \left[ -d_{1} \left( 1 + C_{i,j-1/2,k}^{s,y} \right) \right] \nonumber \\
& + & E_{r  }^{n+1}(i-1,j,k) \left[ -d_{1} \left( 1 + C_{i-1/2,j,k}^{s,x} \right) f_{xx}(i-1,j,k)^{1/2} \right] \nonumber \\
& + & F_{r,x}^{n+1}(i-1,j,k) \left[ -d_{1} \left( 1 + C_{i-1/2,j,k}^{s,x} \right) \right] \nonumber \\
& + & E_{r  }^{n+1}(i,j,k) \left[ 1 
+ d_{1} \left( 1 + C_{i+1/2,j,k}^{s,x} \right) f_{xx}(i,j,k)^{1/2} + d_{1} \left( 1 - C_{i-1/2,j,k}^{s,x} \right) f_{xx}(i,j,k)^{1/2} \right. \nonumber \\
& ~ & ~~~~~~~~~~~~~~~~~~~~ \left. + ~ d_{1} \left( 1 + C_{i,j+1/2,k}^{s,y} \right) f_{yy}(i,j,k)^{1/2} + d_{1} \left( 1 - C_{i,j-1/2,k}^{s,y} \right) f_{yy}(i,j,k)^{1/2} \right. \nonumber \\
& ~ & ~~~~~~~~~~~~~~~~~~~~ \left. + ~ d_{1} \left( 1 + C_{i,j,k+1/2}^{s,z} \right) f_{zz}(i,j,k)^{1/2} + d_{1} \left( 1 - C_{i,j,k-1/2}^{s,z} \right) f_{zz}(i,j,k)^{1/2} + d_{2} \sigma_{a} \right. \nonumber \\
& ~ & ~~~~~~~~~~~~~~~~~~~~ \left. + ~ \frac{ d_{2} (\sigma_{a}-\sigma_{s}) m_{x} }{ \rho^2 \mathbb{C}^2 } \left( m_{x} + m_{x} f_{xx}(i,j,k) + m_{y} f_{yx}(i,j,k) + m_{z} f_{zx}(i,j,k) \right) \right. \nonumber \\
& ~ & ~~~~~~~~~~~~~~~~~~~~ \left. + ~ \frac{ d_{2} (\sigma_{a}-\sigma_{s}) m_{y} }{ \rho^2 \mathbb{C}^2 } \left( m_{y} + m_{x} f_{xy}(i,j,k) + m_{y} f_{yy}(i,j,k) + m_{z} f_{zy}(i,j,k) \right) \right. \nonumber \\
& ~ & ~~~~~~~~~~~~~~~~~~~~ \left. + ~ \frac{ d_{2} (\sigma_{a}-\sigma_{s}) m_{z} }{ \rho^2 \mathbb{C}^2 } \left( m_{z} + m_{x} f_{xz}(i,j,k) + m_{y} f_{yz}(i,j,k) + m_{z} f_{zz}(i,j,k) \right) \right] \nonumber \\
& + & F_{r,x}^{n+1}(i,j,k) \left[ d_{1} \left( 1 + C_{i+1/2,j,k}^{s,x} \right) - d_{1} \left( 1 - C_{i-1/2,j,k}^{s,x} \right) - \frac{ d_{2} (\sigma_{a}-\sigma_{s}) m_{x} }{ \rho \mathbb{C} } \right] \nonumber \\
& + & F_{r,y}^{n+1}(i,j,k) \left[ d_{1} \left( 1 + C_{i,j+1/2,k}^{s,y} \right) - d_{1} \left( 1 - C_{i,j-1/2,k}^{s,y} \right) - \frac{ d_{2} (\sigma_{a}-\sigma_{s}) m_{y} }{ \rho \mathbb{C} } \right] \nonumber \\
& + & F_{r,z}^{n+1}(i,j,k) \left[ d_{1} \left( 1 + C_{i,j,k+1/2}^{s,z} \right) - d_{1} \left( 1 - C_{i,j,k-1/2}^{s,z} \right) - \frac{ d_{2} (\sigma_{a}-\sigma_{s}) m_{z} }{ \rho \mathbb{C} } \right] \nonumber \\
& + & E_{r  }^{n+1}(i+1,j,k) \left[ -d_{1} \left( 1 - C_{i+1/2,j,k}^{s,x} \right) f_{xx}(i+1,j,k)^{1/2} \right] \nonumber \\
& + & F_{r,x}^{n+1}(i+1,j,k) \left[  d_{1} \left( 1 - C_{i+1/2,j,k}^{s,x} \right) \right] \nonumber \\
& + & E_{r  }^{n+1}(i,j+1,k) \left[ -d_{1} \left( 1 - C_{i,j+1/2,k}^{s,y} \right) f_{yy}(i,j+1,k)^{1/2} \right] \nonumber \\
& + & F_{r,y}^{n+1}(i,j+1,k) \left[  d_{1} \left( 1 - C_{i,j+1/2,k}^{s,y} \right) \right] \nonumber \\
& + & E_{r  }^{n+1}(i,j,k+1) \left[ -d_{1} \left( 1 - C_{i,j,k+1/2}^{s,z} \right) f_{zz}(i,j,k+1)^{1/2} \right] \nonumber \\
& + & F_{r,z}^{n+1}(i,j,k+1) \left[  d_{1} \left( 1 - C_{i,j,k+1/2}^{s,z} \right) \right] = E_{r}^{n}(i,j,k) + d_{2} \sigma_{a} T^4 \nonumber , 
\end{eqnarray} 
\begin{eqnarray}
& ~ & E_{r  }^{n+1}(i,j,k-1) \left[ -d_{1} \left( 1 + C_{i,j,k-1/2}^{s,z} \right) f_{xz}(i,j,k-1) \right] \\
& + & F_{r,x}^{n+1}(i,j,k-1) \left[ -d_{1} \left( 1 + C_{i,j,k-1/2}^{s,z} \right) f_{zz}(i,j,k-1)^{1/2} \right] \nonumber \\
& + & E_{r  }^{n+1}(i,j-1,k) \left[ -d_{1} \left( 1 + C_{i,j-1/2,k}^{s,y} \right) f_{xy}(i,j-1,k) \right] \nonumber \\
& + & F_{r,x}^{n+1}(i,j-1,k) \left[ -d_{1} \left( 1 + C_{i,j-1/2,k}^{s,y} \right) f_{yy}(i,j-1,k)^{1/2} \right] \nonumber \\
& + & E_{r  }^{n+1}(i-1,j,k) \left[ -d_{1} \left( 1 + C_{i-1/2,j,k}^{s,x} \right) f_{xx}(i-1,j,k) \right] \nonumber \\
& + & F_{r,x}^{n+1}(i-1,j,k) \left[ -d_{1} \left( 1 + C_{i-1/2,j,k}^{s,x} \right) f_{xx}(i-1,j,k)^{1/2} \right] \nonumber \\
& + & E_{r  }^{n+1}(i,j,k) \left[ ~~~ d_{1} \left( 1 + C_{i+1/2,j,k}^{s,x} \right) f_{xx}(i,j,k) - d_{1} \left( 1 - C_{i-1/2,j,k}^{s,x} \right) f_{xx}(i,j,k) \right. \nonumber \\
& ~ & ~~~~~~~~~~~~~~~~~~ \left. + ~ d_{1} \left( 1 + C_{i,j+1/2,k}^{s,y} \right) f_{xy}(i,j,k) - d_{1} \left( 1 - C_{i,j-1/2,k}^{s,y} \right) f_{xy}(i,j,k) \right. \nonumber \\
& ~ & ~~~~~~~~~~~~~~~~~~ \left. + ~ d_{1} \left( 1 + C_{i,j,k+1/2}^{s,z} \right) f_{xz}(i,j,k) - d_{1} \left( 1 - C_{i,j,k-1/2}^{s,z} \right) f_{xz}(i,j,k) \right. \nonumber \\
& ~ & ~~~~~~~~~~~~~~~~~~ \left. - ~ \frac{ d_{2} \sigma_{t} }{ \rho \mathbb{C} } \left( m_{x} + m_{x} f_{xx}(i,j,k) + m_{y} f_{yx}(i,j,k) + m_{z} f_{zx}(i,j,k) \right) + \frac{ d_{2} \sigma_{a} m_{x} }{ \rho^{n} \mathbb{C} } \right] \nonumber \\
& + & F_{r,x}^{n+1}(i,j,k) \left[1 + d_{1} \left( 1 + C_{i+1/2,j,k}^{s,x} \right) f_{xx}(i,j,k)^{1/2} + d_{1} \left( 1 - C_{i-1/2,j,k}^{s,x} \right) f_{xx}(i,j,k)^{1/2} \right. \nonumber \\
& ~ & ~~~~~~~~~~~~~~~~~~~~ \left. + ~ d_{1} \left( 1 + C_{i,j+1/2,k}^{s,y} \right) f_{yy}(i,j,k)^{1/2} + d_{1} \left( 1 - C_{i,j-1/2,k}^{s,y} \right) f_{yy}(i,j,k)^{1/2} \right. \nonumber \\
& ~ & ~~~~~~~~~~~~~~~~~~~~ \left. + ~ d_{1} \left( 1 + C_{i,j,k+1/2}^{s,z} \right) f_{zz}(i,j,k)^{1/2} + d_{1} \left( 1 - C_{i,j,k-1/2}^{s,z} \right) f_{zz}(i,j,k)^{1/2} + d_{2} \sigma_{t} \right] \nonumber \\
& + & E_{r  }^{n+1}(i+1,j,k) \left[  d_{1} \left( 1 - C_{i+1/2,j,k}^{s,x} \right) f_{xx}(i+1,j,k) \right] \nonumber \\
& + & F_{r,x}^{n+1}(i+1,j,k) \left[ -d_{1} \left( 1 - C_{i+1/2,j,k}^{s,x} \right) f_{xx}(i+1,j,k)^{1/2} \right] \nonumber \\
& + & E_{r  }^{n+1}(i,j+1,k) \left[  d_{1} \left( 1 - C_{i,j+1/2,k}^{s,y} \right) f_{xy}(i,j+1,k) \right] \nonumber \\
& + & F_{r,x}^{n+1}(i,j+1,k) \left[ -d_{1} \left( 1 - C_{i,j+1/2,k}^{s,y} \right) f_{yy}(i,j+1,k)^{1/2} \right] \nonumber \\
& + & E_{r  }^{n+1}(i,j,k+1) \left[  d_{1} \left( 1 - C_{i,j,k+1/2}^{s,z} \right) f_{xz}(i,j,k+1) \right] \nonumber \\
& + & F_{r,x}^{n+1}(i,j,k+1) \left[ -d_{1} \left( 1 - C_{i,j,k+1/2}^{s,z} \right) f_{zz}(i,j,k+1)^{1/2} \right] = F_{r,x}^{n}(i,j,k) + \frac{ d_{2} \sigma_{a} m_{x} T^4 }{ \rho \mathbb{C} } \nonumber , 
\end{eqnarray} 
\begin{eqnarray}
& ~ & E_{r  }^{n+1}(i,j,k-1) \left[ -d_{1} \left( 1 + C_{i,j,k-1/2}^{s,z} \right) f_{yz}(i,j,k-1) \right] \\
& + & F_{r,y}^{n+1}(i,j,k-1) \left[ -d_{1} \left( 1 + C_{i,j,k-1/2}^{s,z} \right) f_{zz}(i,j,k-1)^{1/2} \right] \nonumber \\
& + & E_{r  }^{n+1}(i,j-1,k) \left[ -d_{1} \left( 1 + C_{i,j-1/2,k}^{s,y} \right) f_{yy}(i,j-1,k) \right] \nonumber \\
& + & F_{r,y}^{n+1}(i,j-1,k) \left[ -d_{1} \left( 1 + C_{i,j-1/2,k}^{s,y} \right) f_{yy}(i,j-1,k)^{1/2} \right] \nonumber \\
& + & E_{r  }^{n+1}(i-1,j,k) \left[ -d_{1} \left( 1 + C_{i-1/2,j,k}^{s,x} \right) f_{yx}(i-1,j,k) \right] \nonumber \\
& + & F_{r,y}^{n+1}(i-1,j,k) \left[ -d_{1} \left( 1 + C_{i-1/2,j,k}^{s,x} \right) f_{xx}(i-1,j,k)^{1/2} \right] \nonumber \\
& + & E_{r  }^{n+1}(i,j,k) \left[ ~~~ d_{1} \left( 1 + C_{i+1/2,j,k}^{s,x} \right) f_{yx}(i,j,k) - d_{1} \left( 1 - C_{i-1/2,j,k}^{s,x} \right) f_{yx}(i,j,k) \right. \nonumber \\
& ~ & ~~~~~~~~~~~~~~~~~~ \left. + ~ d_{1} \left( 1 + C_{i,j+1/2,k}^{s,y} \right) f_{yy}(i,j,k) - d_{1} \left( 1 - C_{i,j-1/2,k}^{s,y} \right) f_{yy}(i,j,k) \right. \nonumber \\
& ~ & ~~~~~~~~~~~~~~~~~~ \left. + ~ d_{1} \left( 1 + C_{i,j,k+1/2}^{s,z} \right) f_{yz}(i,j,k) - d_{1} \left( 1 - C_{i,j,k-1/2}^{s,z} \right) f_{yz}(i,j,k) \right. \nonumber \\
& ~ & ~~~~~~~~~~~~~~~~~~ \left. - ~ \frac{ d_{2} \sigma_{t} }{ \rho \mathbb{C} } \left( m_{y} + m_{x} f_{xy}(i,j,k) + m_{y} f_{yy}(i,j,k) + m_{z} f_{zy}(i,j,k) \right) + \frac{ d_{2} \sigma_{a} m_{y} }{ \rho \mathbb{C} } \right] \nonumber \\
& + & F_{r,y}^{n+1}(i,j,k) \left[1 + d_{1} \left( 1 + C_{i+1/2,j,k}^{s,x} \right) f_{xx}(i,j,k)^{1/2} + d_{1} \left( 1 - C_{i-1/2,j,k}^{s,x} \right) f_{xx}(i,j,k)^{1/2} \right. \nonumber \\
& ~ & ~~~~~~~~~~~~~~~~~~~~ \left. + ~ d_{1} \left( 1 + C_{i,j+1/2,k}^{s,y} \right) f_{yy}(i,j,k)^{1/2} + d_{1} \left( 1 - C_{i,j-1/2,k}^{s,y} \right) f_{yy}(i,j,k)^{1/2} \right. \nonumber \\
& ~ & ~~~~~~~~~~~~~~~~~~~~ \left. + ~ d_{1} \left( 1 + C_{i,j,k+1/2}^{s,z} \right) f_{zz}(i,j,k)^{1/2} + d_{1} \left( 1 - C_{i,j,k-1/2}^{s,z} \right) f_{zz}(i,j,k)^{1/2} + d_{2} \sigma_{t} \right] \nonumber \\
& + & E_{r  }^{n+1}(i+1,j,k) \left[  d_{1} \left( 1 - C_{i+1/2,j,k}^{s,x} \right) f_{yx}(i+1,j,k) \right] \nonumber \\
& + & F_{r,y}^{n+1}(i+1,j,k) \left[ -d_{1} \left( 1 - C_{i+1/2,j,k}^{s,x} \right) f_{xx}(i+1,j,k)^{1/2} \right] \nonumber \\
& + & E_{r  }^{n+1}(i,j+1,k) \left[  d_{1} \left( 1 - C_{i,j+1/2,k}^{s,y} \right) f_{yy}(i,j+1,k) \right] \nonumber \\
& + & F_{r,y}^{n+1}(i,j+1,k) \left[ -d_{1} \left( 1 - C_{i,j+1/2,k}^{s,y} \right) f_{yy}(i,j+1,k)^{1/2} \right] \nonumber \\
& + & E_{r  }^{n+1}(i,j,k+1) \left[  d_{1} \left( 1 - C_{i,j,k+1/2}^{s,z} \right) f_{yz}(i,j,k+1) \right] \nonumber \\
& + & F_{r,y}^{n+1}(i,j,k+1) \left[ -d_{1} \left( 1 - C_{i,j,k+1/2}^{s,z} \right) f_{zz}(i,j,k+1)^{1/2} \right] = F_{r,y}^{n}(i,j,k) + \frac{ d_{2} \sigma_{a} m_{y} T^4 }{ \rho \mathbb{C} } \nonumber , 
\end{eqnarray} 
\begin{eqnarray}
& ~ & E_{r  }^{n+1}(i,j,k-1) \left[ -d_{1} \left( 1 + C_{i,j,k-1/2}^{s,z} \right) f_{zz}(i,j,k-1) \right] \\
& + & F_{r,z}^{n+1}(i,j,k-1) \left[ -d_{1} \left( 1 + C_{i,j,k-1/2}^{s,z} \right) f_{zz}(i,j,k-1)^{1/2} \right] \nonumber \\
& + & E_{r  }^{n+1}(i,j-1,k) \left[ -d_{1} \left( 1 + C_{i,j-1/2,k}^{s,y} \right) f_{zy}(i,j-1,k) \right] \nonumber \\
& + & F_{r,z}^{n+1}(i,j-1,k) \left[ -d_{1} \left( 1 + C_{i,j-1/2,k}^{s,y} \right) f_{yy}(i,j-1,k)^{1/2} \right] \nonumber \\
& + & E_{r  }^{n+1}(i-1,j,k) \left[ -d_{1} \left( 1 + C_{i-1/2,j,k}^{s,x} \right) f_{zx}(i-1,j,k) \right] \nonumber \\
& + & F_{r,z}^{n+1}(i-1,j,k) \left[ -d_{1} \left( 1 + C_{i-1/2,j,k}^{s,x} \right) f_{xx}(i-1,j,k)^{1/2} \right] \nonumber \\
& + & E_{r  }^{n+1}(i,j,k) \left[ ~~~ d_{1} \left( 1 + C_{i+1/2,j,k}^{s,x} \right) f_{zx}(i,j,k) - d_{1} \left( 1 - C_{i-1/2,j,k}^{s,x} \right) f_{zx}(i,j,k) \right. \nonumber \\
& ~ & ~~~~~~~~~~~~~~~~~~ \left. + ~ d_{1} \left( 1 + C_{i,j+1/2,k}^{s,y} \right) f_{zy}(i,j,k) - d_{1} \left( 1 - C_{i,j-1/2,k}^{s,y} \right) f_{zy}(i,j,k) \right. \nonumber \\
& ~ & ~~~~~~~~~~~~~~~~~~ \left. + ~ d_{1} \left( 1 + C_{i,j,k+1/2}^{s,z} \right) f_{zz}(i,j,k) - d_{1} \left( 1 - C_{i,j,k-1/2}^{s,z} \right) f_{zz}(i,j,k) \right. \nonumber \\
& ~ & ~~~~~~~~~~~~~~~~~~ \left. - ~ \frac{ d_{2} \sigma_{t} }{ \rho \mathbb{C} } \left( m_{z} + m_{x} f_{xz}(i,j,k) + m_{y} f_{yz}(i,j,k) + m_{z} f_{zz}(i,j,k) \right) + \frac{ d_{2} \sigma_{a} m_{z} }{ \rho \mathbb{C} } \right] \nonumber \\
& + & F_{r,z}^{n+1}(i,j,k) \left[1 + d_{1} \left( 1 + C_{i+1/2,j,k}^{s,x} \right) f_{xx}(i,j,k)^{1/2} + d_{1} \left( 1 - C_{i-1/2,j,k}^{s,x} \right) f_{xx}(i,j,k)^{1/2} \right. \nonumber \\
& ~ & ~~~~~~~~~~~~~~~~~~~~ \left. + ~ d_{1} \left( 1 + C_{i,j+1/2,k}^{s,y} \right) f_{yy}(i,j,k)^{1/2} + d_{1} \left( 1 - C_{i,j-1/2,k}^{s,y} \right) f_{yy}(i,j,k)^{1/2} \right. \nonumber \\
& ~ & ~~~~~~~~~~~~~~~~~~~~ \left. + ~ d_{1} \left( 1 + C_{i,j,k+1/2}^{s,z} \right) f_{zz}(i,j,k)^{1/2} + d_{1} \left( 1 - C_{i,j,k-1/2}^{s,z} \right) f_{zz}(i,j,k)^{1/2} + d_{2} \sigma_{t} \right] \nonumber \\
& + & E_{r  }^{n+1}(i+1,j,k) \left[  d_{1} \left( 1 - C_{i+1/2,j,k}^{s,x} \right) f_{zx}(i+1,j,k) \right] \nonumber \\
& + & F_{r,z}^{n+1}(i+1,j,k) \left[ -d_{1} \left( 1 - C_{i+1/2,j,k}^{s,x} \right) f_{xx}(i+1,j,k)^{1/2} \right] \nonumber \\
& + & E_{r  }^{n+1}(i,j+1,k) \left[  d_{1} \left( 1 - C_{i,j+1/2,k}^{s,y} \right) f_{zy}(i,j+1,k) \right] \nonumber \\
& + & F_{r,z}^{n+1}(i,j+1,k) \left[ -d_{1} \left( 1 - C_{i,j+1/2,k}^{s,y} \right) f_{yy}(i,j+1,k)^{1/2} \right] \nonumber \\
& + & E_{r  }^{n+1}(i,j,k+1) \left[  d_{1} \left( 1 - C_{i,j,k+1/2}^{s,z} \right) f_{zz}(i,j,k+1) \right] \nonumber \\
& + & F_{r,z}^{n+1}(i,j,k+1) \left[ -d_{1} \left( 1 - C_{i,j,k+1/2}^{s,z} \right) f_{zz}(i,j,k+1)^{1/2} \right] = F_{r,z}^{n}(i,j,k) + \frac{ d_{2} \sigma_{a} m_{z} T^4 }{ \rho \mathbb{C} } \nonumber , 
\end{eqnarray} }}

\noindent
where $d_{1} = \Delta t \mathbb{C} / 2 \Delta x$, $d_{2} = \Delta t \mathbb{C}$. All of the material quantities $(\rho,m_{x},m_{y},m_{z},E)$ that appear in the above equations have the spatial index $(i,j,k)$ and temporal index $n$. Additionally, the material temperature enters the above equations via the following relation: $T_{i,j,k}^{n} = p_{i,j,k}^{n} / \rho_{i,j,k}^{n} = \left( \gamma - 1 \right) \left( \frac{E_{i,j,k}^{n}}{\rho_{i,j,k}^{n}} - \frac{1}{2} \frac{\left( M_{i,j,k}^{n} \right)^2}{\left( \rho_{i,j,k}^{n} \right)^2} \right)$, where $M^{2} = m_{x}^{2} + m_{y}^{2} + m_{z}^{2}$. To understand how the multidimensional equations fit into the linear algebraic description $Ax = b$, it is important to understand that the algorithm cycles through the indices $(i,j,k)$ in the following manner: \texttt{for(k=1;N;k++)\{ for(j=1;N;j++) \{ for(i=1;N;i++) \{...\}\}\} }. Furthermore, the algorithm cycles through the radiation quantities in the following order: $(E_{r}$,$F_{r,x}$,$F_{r,y}$,$F_{r,z})$. Assuming that $N_{cell}^{x} = N_{cell}^{y} = N_{cell}^{z}$, one expects the matrix $A$ to have dimensions $\dim(A) = 4N^{3} \times 4N^{3}$. Moreover, the solution vector $x$ will contain the following sequence of entries:
\begin{eqnarray}
x & = & \left[ E_{r}^{n+1}(i,j,k), F_{r,x}^{n+1}(i,j,k), F_{r,y}^{n+1}(i,j,k), F_{r,z}^{n+1}(i,j,k), E_{r}^{n+1}(i+1,j,k), \ldots, \right. \nonumber \\
& ~ & \left. ~ E_{r}^{n+1}(i+2,j,k), \ldots, F_{r,z}^{n+1}(N,j,k), E_{r}^{n+1}(1,j+1,k), \ldots, \right. \nonumber \\
& ~ & \left. ~ F_{r,z}^{n+1}(N,N,k), E_{r}^{n+1}(1,1,k+1), \ldots \right]^{T} . \nonumber
\end{eqnarray} 

\noindent
If one assigns the character $\star$ as a placeholder for non-zero values, then one visualizes the structure of $A$ which results from the above difference equations as a block banded matrix:
{\tiny{
\begin{equation}
\left[ \begin{array}{c}
\vdots ~~~~~~~~~~~~~~~~~~~~~~~~~~~~~~~~~~~~~~~~~~~~~~~~~~~~~~~~~~~~~~~~~~~~~~~~~~~~~~~~~~~~~~~ \ddots ~~~~~~~~~~~~~~~~~~~~~~~~~~~~~~~~~~~~~~~~~~~~~~~~~~~~~~~~~~~~~~~~~~~~~~~~~~~~~~~~~~~~~~~ \nonumber \\
..., 0, \star, 0, 0, \star, \leftrightarrow 0 (N^2-N) \leftrightarrow, \star, 0, \star, 0, \leftrightarrow 0 (N-1) \leftrightarrow, \star, \star, 0, 0, \star, \star, \star, \star, \star, \star, 0, 0, \leftrightarrow 0 (N-1) \leftrightarrow, \star, 0, \star, 0, \leftrightarrow 0 (N^2-N) \leftrightarrow, \star, 0, 0, \star, 0, ... \nonumber \\
..., 0, \star, \star, 0, 0, \leftrightarrow 0 (N^2-N) \leftrightarrow, \star, \star, 0, 0, \leftrightarrow 0 (N-1) \leftrightarrow, \star, \star, 0, 0, \star, \star, 0, 0, \star, \star, 0, 0, \leftrightarrow 0 (N-1) \leftrightarrow, \star, \star, 0, 0, \leftrightarrow 0 (N^2-N) \leftrightarrow, \star, \star, 0, 0, 0, ... \nonumber \\
..., 0, \star, 0, \star, 0, \leftrightarrow 0 (N^2-N) \leftrightarrow, \star, 0, \star, 0, \leftrightarrow 0 (N-1) \leftrightarrow, \star, 0, \star, 0, \star, 0, \star, 0, \star, 0, \star, 0, \leftrightarrow 0 (N-1) \leftrightarrow, \star, 0, \star, 0, \leftrightarrow 0 (N^2-N) \leftrightarrow, \star, 0, \star, 0, 0, ... \nonumber \\
..., 0, \star, 0, 0, \star, \leftrightarrow 0 (N^2-N) \leftrightarrow, \star, 0, 0, \star, \leftrightarrow 0 (N-1) \leftrightarrow, \star, 0, 0, \star, \star, 0, 0, \star, \star, 0, 0, \star, \leftrightarrow 0 (N-1) \leftrightarrow, \star, 0, 0, \star, \leftrightarrow 0 (N^2-N) \leftrightarrow, \star, 0, 0, \star, 0, ... \nonumber \\
\vdots ~~~~~~~~~~~~~~~~~~~~~~~~~~~~~~~~~~~~~~~~~~~~~~~~~~~~~~~~~~~~~~~~~~~~~~~~~~~~~~~~~~~~~~~ \ddots ~~~~~~~~~~~~~~~~~~~~~~~~~~~~~~~~~~~~~~~~~~~~~~~~~~~~~~~~~~~~~~~~~~~~~~~~~~~~~~~~~~~~~~~ \nonumber 
\end{array} \right] .
\end{equation} }}

\noindent
Since there are no nonlinearities in the radiation quantities for which root finding (e.g., Newton's method) must be implemented, one casts these equations into a sparse matrix format that can be solved with iteration techniques from linear algebra such as the Jacobi, Gauss-Seidel, and multigrid methods as well as others because $A$ exhibits diagonal dominance.


\section{Modified Godunov Predictor Scheme}
\noindent
Given that the radiation quantities $(E_{r}$,$F_{r,x}$,$F_{r,y}$,$F_{r,z})$ are at time $t_{n+1}$, one computes the flux divergence $( (\nabla \cdot F^{m})^{n+1/2}$,$(\nabla \cdot G^{m})^{n+1/2}$,$(\nabla \cdot H^{m})^{n+1/2} )$ for the material quantities $(\rho, m_{x}, m_{y}, m_{z}, E)$ that are at time $t_{n}$. Following the analysis of \cite{mc2007, sekora2009, sekora2010, treb2005}, one applies Duhamel's principle to the quasi-linear system of balance laws in Equations \ref{eq:cons_law_quasi}-\ref{eq:cons_law_jacobian_az} for only the material components. This technique defines the following system that locally includes in space and time the effects of the stiff source terms on the hyperbolic structure:
\begin{equation}
\frac{D U^{m}_{\textrm{eff}}}{Dt} = \mathcal{I}(\eta) \left( - A^{m}_{x,L} \partdif{U^{m}}{x} - A^{m}_{y,L} \partdif{U^{m}}{y} - A^{m}_{z,L} \partdif{U^{m}}{z} + S^{m}(U^{m,n},U^{r,n+1}) \right) ,
\end{equation}

\noindent 
where $\frac{DU^{m}}{Dt} = \partdif{U^{m}}{t} + u \partdif{U^{m}}{x} + v \partdif{U^{m}}{y} + w \partdif{U^{m}}{z}$ is the total derivative. $\mathcal{I}$ is a propagation operator that projects the dynamics of the stiff source terms onto the hyperbolic structure. $A^{m}_{x,L} = A_{x}^{m} - u I$, $A^{m}_{y,L} = A_{y}^{m} - v I$, and $A^{m}_{z,L} = A_{z}^{m} - w I$ are Jacobians associated with Lagrangian trajectories in the $x$, $y$, and $z$ directions, respectively. Since the predictor scheme is a first-order accurate step in a second-order accurate predictor-corrector method, one chooses $\eta = \Delta t / 2$ and the effective balance law becomes:
\begin{equation} 
\partdif{U^{m}}{t} + A^{m}_{x,\textrm{eff}} \partdif{U^{m}}{x} + A^{m}_{y,\textrm{eff}} \partdif{U^{m}}{y} + A^{m}_{z,\textrm{eff}} \partdif{U^{m}}{z} = \mathcal{I}(\Delta t / 2) S^{m}(U^{m,n},U^{r,n+1}) ,
\end{equation}

\noindent 
where $A^{m}_{x,\textrm{eff}} = \mathcal{I}(\Delta t / 2) A^{m}_{x,L} + u I$, $A^{m}_{y,\textrm{eff}} = \mathcal{I}(\Delta t / 2) A^{m}_{y,L} + v I$, and $A^{m}_{z,\textrm{eff}} = \mathcal{I}(\Delta t / 2) A^{m}_{z,L} + w I$.

\subsection{Applying the Modified Godunov Predictor Scheme}
\noindent
If one considers only the material component in Equations \ref{eq:rh1}-\ref{eq:rh3}, then the variables, fluxes, and source terms are:
\begin{equation}
U = 
\left( \begin{array}{c} \rho    \\
                        m_{x}   \\
                        m_{y}   \\
                        m_{z}   \\
                        E       \end{array} \right) , ~~~~
F(U) = 
\left( \begin{array}{c} m_{x} \\
                        \frac{m_{x}^2}{\rho} + p  \\
                        \frac{m_{x} m_{y}}{\rho}  \\
                        \frac{m_{x} m_{z}}{\rho}  \\
                        (E + p)\frac{m_{x}}{\rho} \end{array} \right) , ~~~~
G(U) = 
\left( \begin{array}{c} m_{y} \\
                        \frac{m_{y} m_{x}}{\rho}  \\
                        \frac{m_{y}^2}{\rho} + p  \\
                        \frac{m_{y} m_{z}}{\rho}  \\
                        (E + p)\frac{m_{y}}{\rho} \end{array} \right) , \nonumber
\end{equation}
\begin{equation}
H(U) = 
\left( \begin{array}{c} m_{z} \\
                        \frac{m_{z} m_{x}}{\rho}  \\
                        \frac{m_{z} m_{y}}{\rho}  \\
                        \frac{m_{z}^2}{\rho} + p  \\
                        (E + p)\frac{m_{z}}{\rho} \end{array} \right) , ~~~~
S^{m}(U) = 
\left( \begin{array}{c} 0 \\
                        -\mathbb{P} S^{F_x} \\
                        -\mathbb{P} S^{F_y} \\
                        -\mathbb{P} S^{F_z} \\
                        -\mathbb{P} \mathbb{C} S^E \end{array} \right) . \nonumber
\end{equation}

\noindent
In order to compute $\mathcal{I}$, one first computes $\nabla_{U^{m}} S^{m}(U)$. Therefore:
\begin{equation}
\nabla_{U^{m}} S^{m}(U) = \left( \begin{array}{ccccc}
0 & 0 & 0 & 0 & 0 \\
-\mathbb{P} S^{F_{x}}_{\rho} & -\mathbb{P} S^{F_{x}}_{m_{x}} & -\mathbb{P} S^{F_{x}}_{m_{y}} & -\mathbb{P} S^{F_{x}}_{m_{z}} & -\mathbb{P} S^{F_{x}}_{E} \\
-\mathbb{P} S^{F_{y}}_{\rho} & -\mathbb{P} S^{F_{y}}_{m_{x}} & -\mathbb{P} S^{F_{y}}_{m_{y}} & -\mathbb{P} S^{F_{y}}_{m_{z}} & -\mathbb{P} S^{F_{y}}_{E} \\
-\mathbb{P} S^{F_{z}}_{\rho} & -\mathbb{P} S^{F_{z}}_{m_{x}} & -\mathbb{P} S^{F_{z}}_{m_{y}} & -\mathbb{P} S^{F_{z}}_{m_{z}} & -\mathbb{P} S^{F_{z}}_{E} \\
-\mathbb{P} \mathbb{C} S^{E}_{\rho} & -\mathbb{P} \mathbb{C} S^{E}_{m_{x}} & -\mathbb{P} \mathbb{C} S^{E}_{m_{y}} & -\mathbb{P} \mathbb{C} S^{E}_{m_{z}} & -\mathbb{P} \mathbb{C} S^{E}_{E} 
\end{array} \right) \label{eq:dSodU} ,
\end{equation}

\noindent
where the partial derivatives are presented in Appendix 1.

\subsubsection{Simplifying $\nabla_{U^{m}} S^{m}(U)$}
\noindent
In its current form, $\nabla_{U^{m}} S^{m}(U)$ in Equation \ref{eq:dSodU} leads to a propagation operator $\mathcal{I}$ that is difficult to work with algebraically. By inspection, the material momentum source terms are not the dominant factors defining the stiffness associated with the matter-radiation coupling, such that $-\mathbb{P} \mathbf{S}^{\mathbf{F}} < \mathcal{O}(1)$ even in the strong equilibrium diffusion limit. Additionally, one finds that the derivative of the material momentum source term with respect to the variables $U^{m}$ has the following magnitude $-\mathbb{P} \mathbf{S}^{\mathbf{F}}_{U^{m}} < \mathcal{O}(1)$. Therefore, $-\mathbb{P} \mathbf{S}^{\mathbf{F}}$ can be included as a body force (e.g., gravity).  \\

\noindent
It is the material energy source term $-\mathbb{P} \mathbb{C} S^{E}$ that defines the stiffness associated with the problem. By inspection of the contributing terms, $-\mathbb{P} \mathbb{C} S^{E} < \mathcal{O}(\mathbb{C})$ in the strong equilibrium diffusion limit. Additionally, one finds that the derivative of the material energy source term with respect to the variables $U^{m}$ has the following magnitude $-\mathbb{P} \mathbb{C} S^{E}_{U^{m}} < \mathcal{O}(\mathbb{C}^{2})$. Therefore, one only needs to use $-\mathbb{P} \mathbb{C} S^{E}$ to define $\nabla_{U^{m}} S^{m}(U)$, such that:
\begin{equation}
\nabla_{U^{m}} S^{m}(U) = \left( \begin{array}{ccccc}
0 & 0 & 0 & 0 & 0 \\
0 & 0 & 0 & 0 & 0 \\
0 & 0 & 0 & 0 & 0 \\
0 & 0 & 0 & 0 & 0 \\
-\mathbb{P} \mathbb{C} S^{E}_{\rho} & -\mathbb{P} \mathbb{C} S^{E}_{m_{x}} & -\mathbb{P} \mathbb{C} S^{E}_{m_{y}} & -\mathbb{P} \mathbb{C} S^{E}_{m_{z}} & -\mathbb{P} \mathbb{C} S^{E}_{E} 
\end{array} \right) \label{eq:dSodU_simp} .
\end{equation}

\noindent
$\nabla_{U^{m}} S^{m}(U)$ is further simplified by examining $S^{E}_{\rho}$, $S^{E}_{m_{x}}$, $S^{E}_{m_{y}}$, $S^{E}_{m_{z}}$, and $S^{E}_{E}$ in the equilibrium diffusion limit and neglecting terms that have magnitudes of or less than $\mathcal{O}(\mathbb{C})$. Therefore:
\begin{eqnarray}
S^{E}_{\rho} & = & 4 \sigma_{a} T^{3} \left( \gamma - 1 \right) \left( \frac{-E}{\rho^{2}} + \frac{M^{2}}{\rho^{3}} \right) \label{simp:partfe1} \\ 
S^{E}_{m_{x}} & = & 4 \sigma_{a} T^{3} \left( \gamma - 1 \right) \left( \frac{-m_{x}}{\rho^{2}} \right) \label{simp:partfe2} \\
S^{E}_{m_{y}} & = & 4 \sigma_{a} T^{3} \left( \gamma - 1 \right) \left( \frac{-m_{y}}{\rho^{2}} \right) \label{simp:partfe3} \\
S^{E}_{m_{z}} & = & 4 \sigma_{a} T^{3} \left( \gamma - 1 \right) \left( \frac{-m_{z}}{\rho^{2}} \right) \label{simp:partfe4} \\
S^{E}_{E}     & = & 4 \sigma_{a} T^{3} \left( \gamma - 1 \right) \left( \frac{1}{\rho} \right) \label{simp:partfe5} .
\end{eqnarray}

\noindent
It is important to note that these partial derivatives have the same stiff magnitude $4 \sigma_{a} T^{3} \left( \gamma - 1 \right)$. This insight simplifies algebraic manipulation. \\

\noindent
If $\nabla_{U^{m}} S^{m}(U)$ is diagonalizable, then $\nabla_{U^{m}} S^{m}(U) = R D R^{-1}$. Here, $D = \textrm{diag}(0,0,0,0,-\mathbb{P} \mathbb{C} S^{E}_{E})$ and $R$ is a matrix whose columns are the right eigenvectors. Below, one sees how the stiff magnitudes cancel out:
\begin{equation}
R = \left( \begin{array}{ccccc}
\frac{-S^{E}_{E}}{S^{E}_{\rho}} & \frac{-S^{E}_{m_{z}}}{S^{E}_{\rho}} & \frac{-S^{E}_{m_{y}}}{S^{E}_{\rho}} & \frac{-S^{E}_{m_{x}}}{S^{E}_{\rho}} & 0 \\
0 & 0 & 0 & 1 & 0 \\
0 & 0 & 1 & 0 & 0 \\
0 & 1 & 0 & 0 & 0 \\
1 & 0 & 0 & 0 & 1 \end{array} \right) , ~~
R^{-1} = \left( \begin{array}{ccccc}
\frac{-S^{E}_{\rho}}{S^{E}_{E}} & \frac{-S^{E}_{m_{x}}}{S^{E}_{E}} & \frac{-S^{E}_{m_{y}}}{S^{E}_{E}} & \frac{-S^{E}_{m_{z}}}{S^{E}_{E}} & 0 \\
0 & 0 & 0 & 1 & 0 \\
0 & 0 & 1 & 0 & 0 \\
0 & 1 & 0 & 0 & 0 \\
\frac{S^{E}_{\rho}}{S^{E}_{E}} & \frac{S^{E}_{m_{x}}}{S^{E}_{E}} & \frac{S^{E}_{m_{y}}}{S^{E}_{E}} & \frac{S^{E}_{m_{z}}}{S^{E}_{E}} & 1 \end{array} \right) .
\end{equation}

\subsubsection{Propagation Operator $\mathcal{I}$}
\noindent
Since one is considering a modified Godunov scheme with a predictor step of $\Delta t/2$ \cite{mc2007}:
\begin{eqnarray}
\mathcal{I} \left( \frac{\Delta t}{2} \right) & = & \frac{1}{\Delta t/2} \int^{\Delta t/2}_{0} e^{\tau \nabla_{U^{m}} S^{m}(U)} d\tau \\
& = & \left( \begin{array}{ccccc}
            1 & 0 & 0 & 0 & 0 \\
            0 & 1 & 0 & 0 & 0 \\
            0 & 0 & 1 & 0 & 0 \\
            0 & 0 & 0 & 1 & 0 \\
            (\alpha-1) \frac{S^{E}_{\rho}}{S^{E}_{E}} & (\alpha-1) \frac{S^{E}_{m_{x}}}{S^{E}_{E}} & (\alpha-1) \frac{S^{E}_{m_{y}}}{S^{E}_{E}} & (\alpha-1) \frac{S^{E}_{m_{z}}}{S^{E}_{E}} & \alpha  
            \end{array} \right) \\
& = & \left( \begin{array}{ccccc}
            1 & 0 & 0 & 0 & 0 \\
            0 & 1 & 0 & 0 & 0 \\
            0 & 0 & 1 & 0 & 0 \\
            0 & 0 & 0 & 1 & 0 \\
            (1-\alpha) \left( \frac{E}{\rho} - \frac{M^2}{\rho^2} \right) & (1-\alpha) \frac{m_{x}}{\rho} & (1-\alpha) \frac{m_{y}}{\rho} & (1-\alpha) \frac{m_{z}}{\rho} & \alpha  
            \end{array} \right) ,
\end{eqnarray} 

\noindent 
where $\alpha = \left( 1 - \exp ( -\mathbb{P} \mathbb{C} S^{E}_{E} \Delta t / 2 ) \right) / \left( \mathbb{P} \mathbb{C} S^{E}_{E} \Delta t / 2 \right)$. Since $S^{E}_{E} \geq 0$ across all relevant parameter space, $0 \leq \alpha \leq 1$. This property is important when considering stability and the subcharacteristic condition which is discussed later in this paper.

\subsection{Effective Material Jacobians - $A^{m}_{x,\textrm{eff}}$, $A^{m}_{y,\textrm{eff}}$, $A^{m}_{z,\textrm{eff}}$}
\noindent
The effects of the stiff source terms on the hyperbolic structure are accounted for by transforming to a moving-mesh (Lagrangian) frame ($A^{m}_{x,L} = A^{m}_{x} - uI$, $A^{m}_{y,L} = A^{m}_{y} - vI$, $A^{m}_{z,L} = A^{m}_{z} - wI$), applying the propagation operator $\mathcal{I}$ to $A^{m}_{L}$, and transforming back to an Eulerian frame, such that the effective material Jacobians ($A^{m}_{x,\textrm{eff}} = \mathcal{I} A^{m}_{x,L} + uI$, $A^{m}_{y,\textrm{eff}} = \mathcal{I} A^{m}_{y,L} + vI$, $A^{m}_{z,\textrm{eff}} = \mathcal{I} A^{m}_{z,L} + wI$) are given by \cite{mc2007}:

{\tiny {\begin{equation}
A^{m}_{x,\textrm{eff}} = \left( \begin{array}{ccccc} 
0                                                 & 1                       & 0             & 0             & 0          \\
\frac{\gamma-1}{2} V^2 - u^2                      & -(\gamma-3) u           & -(\gamma-1) v & -(\gamma-1) w & (\gamma-1) \\
-u v                                              & v                       & u             & 0             & 0          \\
-u w                                              & w                       & 0             & u             & 0          \\
u \left[ \frac{\gamma-1}{2} V^2 - \alpha \tilde{H} - (1-\alpha) \left( \frac{T}{\gamma-1} + \frac{1}{2} V^{2} \right) \right] & -(\gamma-1)u^2 + \alpha \tilde{H} + (1-\alpha) \left( \frac{T}{\gamma-1} + \frac{1}{2} V^{2} \right) & -(\gamma-1)uv & -(\gamma-1)uw & \gamma u
\end{array} \right) , \nonumber
\end{equation} 

\begin{equation}
A^{m}_{y,\textrm{eff}} = \left( \begin{array}{ccccc} 
0                                                 & 0                       & 1             & 0             & 0          \\
-u v                                              & v                       & u             & 0             & 0          \\
\frac{\gamma-1}{2} V^2 - v^2                      & -(\gamma-1) u           & -(\gamma-3) v & -(\gamma-1) w & (\gamma-1) \\
-v w                                              & 0                       & w             & v             & 0          \\
v \left[ \frac{\gamma-1}{2} V^2 - \alpha \tilde{H} - (1-\alpha) \left( \frac{T}{\gamma-1} + \frac{1}{2} V^{2} \right) \right] & -(\gamma-1)uv & -(\gamma-1)v^2 + \alpha \tilde{H} + (1-\alpha) \left( \frac{T}{\gamma-1} + \frac{1}{2} V^{2} \right) & -(\gamma-1)vw & \gamma v   
\end{array} \right), \nonumber 
\end{equation}

\begin{equation}
A^{m}_{z,\textrm{eff}} = \left( \begin{array}{ccccc} 
0                                                 & 0                       & 0             & 1             & 0          \\
-u w                                              & w                       & 0             & u             & 0          \\
-v w                                              & 0                       & w             & v             & 0          \\
\frac{\gamma-1}{2} V^2 - w^2                      & -(\gamma-1) u           & -(\gamma-1) v & -(\gamma-3) w & (\gamma-1) \\
w \left[ \frac{\gamma-1}{2} V^2 - \alpha \tilde{H} - (1-\alpha) \left( \frac{T}{\gamma-1} + \frac{1}{2} V^{2} \right) \right] & -(\gamma-1)uw & -(\gamma-1)vw & -(\gamma-1)w^2 + \alpha \tilde{H} + (1-\alpha) \left( \frac{T}{\gamma-1} + \frac{1}{2} V^{2} \right) & \gamma w   
\end{array} \right) \nonumber . 
\end{equation} }}

\noindent 
These Jacobians have the following eigenvalues: $\lambda^{m}_{x,\textrm{eff},\{-,0,+\}} = \{u-a_{\textrm{eff}}, u, u+a_{\textrm{eff}} \}$, $\lambda^{m}_{y,\textrm{eff},\{-,0,+\}} = \{v-a_{\textrm{eff}}, v, v+a_{\textrm{eff}} \}$, and $\lambda^{m}_{z,\textrm{eff},\{-,0,+\}} = \{w-a_{\textrm{eff}}, w, w+a_{\textrm{eff}} \}$, respectively. Here, the effective sound speed $a_{\textrm{eff}}$ (i.e., the radiation modified sound speed) is:
\begin{eqnarray}
a^{2}_{\textrm{eff}} & = & -\frac{\gamma-1}{2} V^2 + \alpha (\gamma-1) \tilde{H} + (1-\alpha) \left( T + \frac{\gamma-1}{2} V^2 \right)  \\
~                    & = & \alpha \frac{\gamma p}{\rho} + (1-\alpha) T  \\
~                    & = & \left( \alpha (\gamma-1) + 1  \right) \frac{p}{\rho} ,
\end{eqnarray}

\noindent
where $T = p / \rho$ because of the choice of non-dimensionalization. Here, one notices that $\tilde{H}, \left( T + (\gamma-1) V^2/2 \right) \geq 0$ across all relevant parameter space such that the effective sound speed $a_{\textrm{eff}}$ admits the following limits:
\begin{eqnarray}  
-\mathbb{P} \mathbb{C} S^{E}_{E} \rightarrow 0 \Rightarrow \alpha \rightarrow 1 & \Rightarrow & a^{2}_{\textrm{eff}} \rightarrow -\frac{\gamma-1}{2} V^2 + (\gamma-1) \tilde{H} = \frac{\gamma p}{\rho} ~~~~ \textrm{(adiabatic)}  \label{eq:aeff.ad} \\
-\mathbb{P} \mathbb{C} S^{E}_{E} \rightarrow -\infty \Rightarrow \alpha \rightarrow 0 & \Rightarrow & a^{2}_{\textrm{eff}} \rightarrow T = \frac{p}{\rho} ~~~~ \textrm{(isothermal)} . \label{eq:aeff.iso}
\end{eqnarray}

\noindent
When examining Equations \ref{eq:aeff.ad} and \ref{eq:aeff.iso}, one sees that the subcharacteristic condition for material wave speeds is satisfied in each spatial direction, such that \cite{mc2007}:
{\footnotesize {\begin{equation}
\lambda^{m}_{x,-} = u-a_{\textrm{ad}} \leq \lambda^{m}_{x,\textrm{eff},-} = u-a_{\textrm{eff}} \leq \lambda^{m}_{x,0} = \lambda^{m}_{x,\textrm{eff},0} = u \leq \lambda^{m}_{x,\textrm{eff},+} = u+a_{\textrm{eff}} \leq \lambda^{m}_{x,\textrm{eff},+} = u+a_{\textrm{ad}} ,
\end{equation} 
\begin{equation}
\lambda^{m}_{y,-} = v-a_{\textrm{ad}} \leq \lambda^{m}_{y,\textrm{eff},-} = v-a_{\textrm{eff}} \leq \lambda^{m}_{y,0} = \lambda^{m}_{y,\textrm{eff},0} = v \leq \lambda^{m}_{y,\textrm{eff},+} = v+a_{\textrm{eff}} \leq \lambda^{m}_{y,\textrm{eff},+} = v+a_{\textrm{ad}} ,
\end{equation}
\begin{equation}
\lambda^{m}_{z,-} = w-a_{\textrm{ad}} \leq \lambda^{m}_{z,\textrm{eff},-} = w-a_{\textrm{eff}} \leq \lambda^{m}_{z,0} = \lambda^{m}_{z,\textrm{eff},0} = w \leq \lambda^{m}_{z,\textrm{eff},+} = w+a_{\textrm{eff}} \leq \lambda^{m}_{z,\textrm{eff},+} = w+a_{\textrm{ad}} ,
\end{equation} }}

\noindent
This condition is necessary for the stability of the system and guarantees that the numerical solution tends to the solution of the equilibrium equation as the relaxation time tends to zero. Additionally, the structure of the equations remains consistent with respect to classical Godunov methods. Therefore, the CFL CTU time-stepping condition applies. Lastly, the right material eigenvectors $R^{m}_{\{x,y,z\},\textrm{eff}}$ (stored as columns) and left material eigenvectors $L^{m}_{\{x,y,z\},\textrm{eff}}$ (stored as rows) are given in Appendix 2.

\subsection{Computing Left/Right States}
\noindent
In the modified Godunov predictor scheme one uses effective piecewise linear extrapolation to spatially reconstruct material quantities at the left/right sides of cell interfaces. This technique is shown in the following relations for each spatial direction:
\begin{eqnarray}
\tilde{U}_{L,i+1/2,j,k}^{m,n+1/2} = U_{i,j,k}^{m,n}   &+& \frac{1}{2} \left( I - \frac{\Delta t}{\Delta x} A_{x,\textrm{eff}}^{m}(U_{i,j,k}^{m,n}) \right) P^{x}_{\Delta}(\Delta U_{i,j,k}^{m,n}) \\
                                                      &+& \frac{\Delta t}{2} \mathcal{I} \left( \frac{\Delta t}{2} \right) S^{m}(U_{i,j,k}^{m,n},U_{i,j,k}^{r,n+1}) \nonumber , \\
\tilde{U}_{R,i+1/2,j,k}^{m,n+1/2} = U_{i+1,j,k}^{m,n} &-& \frac{1}{2} \left( I + \frac{\Delta t}{\Delta x} A_{x,\textrm{eff}}^{m}(U_{i+1,j,k}^{m,n}) \right) P^{x}_{\Delta}(\Delta U_{i+1,j,k}^{m,n}) \\
                                                      &+& \frac{\Delta t}{2} \mathcal{I} \left( \frac{\Delta t}{2} \right) S^{m}(U_{i+1,j,k}^{m,n},U_{i+1,j,k}^{r,n+1}) \nonumber ,
\end{eqnarray}
\begin{eqnarray}
\tilde{U}_{L,i,j+1/2,k}^{m,n+1/2} = U_{i,j,k}^{m,n}   &+& \frac{1}{2} \left( I - \frac{\Delta t}{\Delta y} A_{y,\textrm{eff}}^{m}(U_{i,j,k}^{m,n}) \right) P^{y}_{\Delta}(\Delta U_{i,j,k}^{m,n}) \\
                                                      &+& \frac{\Delta t}{2} \mathcal{I} \left( \frac{\Delta t}{2} \right) S^{m}(U_{i,j,k}^{m,n},U_{i,j,k}^{r,n+1}) \nonumber , \\
\tilde{U}_{R,i,j+1/2,k}^{m,n+1/2} = U_{i,j+1,k}^{m,n} &-& \frac{1}{2} \left( I + \frac{\Delta t}{\Delta y} A_{y,\textrm{eff}}^{m}(U_{i,j+1,k}^{m,n}) \right) P^{y}_{\Delta}(\Delta U_{i,j+1,k}^{m,n}) \\
                                                      &+& \frac{\Delta t}{2} \mathcal{I} \left( \frac{\Delta t}{2} \right) S^{m}(U_{i,j+1,k}^{m,n},U_{i,j+1,k}^{r,n+1}) \nonumber ,
\end{eqnarray}
\begin{eqnarray}
\tilde{U}_{L,i,j,k+1/2}^{m,n+1/2} = U_{i,j,k}^{m,n}   &+& \frac{1}{2} \left( I - \frac{\Delta t}{\Delta z} A_{z,\textrm{eff}}^{m}(U_{i,j,k}^{m,n}) \right) P^{z}_{\Delta}(\Delta U_{i,j,k}^{m,n}) \\
                                                      &+& \frac{\Delta t}{2} \mathcal{I} \left( \frac{\Delta t}{2} \right) S^{m}(U_{i,j,k}^{m,n},U_{i,j,k}^{r,n+1}) \nonumber , \\
\tilde{U}_{R,i,j,k+1/2}^{m,n+1/2} = U_{i,j,k+1}^{m,n} &-& \frac{1}{2} \left( I + \frac{\Delta t}{\Delta z} A_{z,\textrm{eff}}^{m}(U_{i,j,k+1}^{m,n}) \right) P^{z}_{\Delta}(\Delta U_{i,j,k+1}^{m,n}) \\
                                                      &+& \frac{\Delta t}{2} \mathcal{I} \left( \frac{\Delta t}{2} \right) S^{m}(U_{i,j,k+1}^{m,n},U_{i,j,k+1}^{r,n+1}) \nonumber .
\end{eqnarray}

\noindent
where $P^{\{x,y,z\}}_{\Delta}$ is a slope limiting function used to eliminate spurious oscillations in the $x$, $y$, and $z$ directions, re$y$, and$y$, and$y$, and $z$ directions, re$y$, and$y$, and $z$ directions, re$y$, and $z$ directions, respectively. Slope limiting is performed for each of the material quantities. Although most slope limiters can be used, this algorithm employs the extremum-preserving \cite{colellasekora2008, sekoracolella2009} and traditional van Leer limiters (also referred to as the MUSCL limiter). These techniques can be implemented either componentwise or across characteristic fields. After reconstructing the material quantities in each of the spatial directions, an approximate Riemann solver evaluates the passage of material into and out of each computational cell by using the material states that are to the left/right of the cell interfaces \cite{roe1981}. It is important to emphasize that these flux functions do not directly account for the influence of radiation on the material quantities. Rather, the radiation effects are included via the source terms, propagation operator, and effective material Jacobian.


\section{Corner Transport Upwind Correction}
\noindent
Before advancing the material quantities from time $t_{n}$ to time $t_{n+1}$, one accounts for how the left/right states $( \tilde{U}_{L/R,i+1/2,j,k}^{m}$, $\tilde{U}_{L/R,i,j+1/2,k}^{m}$, $\tilde{U}_{L/R,i,j,k+1/2}^{m} )$ and thus the fluxes $( \tilde{F}_{i+1/2,j,k}^{m}$, $\tilde{G}_{i,j+1/2,k}^{m}$, $\tilde{H}_{i,j,k+1/2}^{m} )$ are affected by transport in the other spatial directions. In particular, one corrects for material propagating across the corners of a computational cell in the following manner \cite{colella1990}:
\begin{eqnarray}
U_{L,i+1/2,j,k}^{m,n+1/2} = \tilde{U}_{L,i+1/2,j,k}^{m,n+1/2} &-& \frac{\Delta t}{2 \Delta y} \left( \tilde{G}_{i,j+1/2,k}^{m} - \tilde{G}_{i,j-1/2,k}^{m} \right) \\
&-& \frac{\Delta t}{2 \Delta z} \left( \tilde{H}_{i,j,k+1/2}^{m} - \tilde{H}_{i,j,k-1/2}^{m} \right) , \nonumber \\
U_{R,i+1/2,j,k}^{m,n+1/2} = \tilde{U}_{R,i+1/2,j,k}^{m,n+1/2} &-& \frac{\Delta t}{2 \Delta y} \left( \tilde{G}_{i+1,j+1/2,k}^{m} - \tilde{G}_{i+1,j-1/2,k}^{m} \right) \\
&-& \frac{\Delta t}{2 \Delta z} \left( \tilde{H}_{i+1,j,k+1/2}^{m} - \tilde{H}_{i+1,j,k-1/2}^{m} \right) , \nonumber 
\end{eqnarray}
\begin{eqnarray}
U_{L,i,j+1/2,k}^{m,n+1/2} = \tilde{U}_{L,i,j+1/2,k}^{m,n+1/2} &-& \frac{\Delta t}{2 \Delta x} \left( \tilde{F}_{i+1/2,j,k}^{m} - \tilde{F}_{i-1/2,j,k}^{m} \right) \\
&-& \frac{\Delta t}{2 \Delta z} \left( \tilde{H}_{i,j,k+1/2}^{m} - \tilde{H}_{i,j,k-1/2}^{m} \right) , \nonumber \\
U_{R,i,j+1/2,k}^{m,n+1/2} = \tilde{U}_{R,i,j+1/2,k}^{m,n+1/2} &-& \frac{\Delta t}{2 \Delta x} \left( \tilde{F}_{i+1/2,j+1,k}^{m} - \tilde{F}_{i-1/2,j+1,k}^{m} \right) \\
&-& \frac{\Delta t}{2 \Delta z} \left( \tilde{H}_{i,j+1,k+1/2}^{m} - \tilde{H}_{i,j+1,k-1/2}^{m} \right) , \nonumber 
\end{eqnarray}
\begin{eqnarray}
U_{L,i,j,k+1/2}^{m,n+1/2} = \tilde{U}_{L,i,j,k+1/2}^{m,n+1/2} &-& \frac{\Delta t}{2 \Delta x} \left( \tilde{F}_{i+1/2,j,k}^{m} - \tilde{F}_{i-1/2,j,k}^{m} \right) \\
&-& \frac{\Delta t}{2 \Delta y} \left( \tilde{G}_{i,j+1/2,k}^{m} - \tilde{G}_{i,j-1/2,k}^{m} \right) , \nonumber \\
U_{R,i,j,k+1/2}^{m,n+1/2} = \tilde{U}_{R,i,j,k+1/2}^{m,n+1/2} &-& \frac{\Delta t}{2 \Delta x} \left( \tilde{F}_{i+1/2,j,k+1}^{m} - \tilde{F}_{i-1/2,j,k+1}^{m} \right) \\
&-& \frac{\Delta t}{2 \Delta y} \left( \tilde{G}_{i,j+1/2,k+1}^{m} - \tilde{G}_{i,j-1/2,k+1}^{m} \right) . \nonumber 
\end{eqnarray}

\noindent
With these corrected left/right material states, one again uses an approximate Riemann solver to evaluate the passage of material.


\section{Modified Godunov Corrector Scheme}
\noindent
The time discretization for the source term is a single-step, second order accurate scheme based on \cite{dutt2000, mc2007, minion2003}. Given the material system of balance laws, one aims for a scheme that has an explicit approach for the flux divergence $(\nabla \cdot F^{m} + \nabla \cdot G^{m} + \nabla \cdot H^{m})$ and an implicit approach for the stiff source term $S^{m}(U)$. At each grid point, one solves the following collection of ordinary differential equations:
\begin{equation}
\frac{dU^{m}}{dt} = S^{m}(U) - ( \nabla \cdot F^{m} )^{n+1/2} - ( \nabla \cdot G^{m} )^{n+1/2} - ( \nabla \cdot H^{m} )^{n+1/2},
\end{equation}

\noindent
where the time-centered flux divergence terms are inputted from the predictor step and taken to be constant valued. Using Picard iteration and the method of deferred corrections, an initial guess for the solution to the collection of ordinary differential equations is:
{\small {
\begin{eqnarray}
\hat{U} = U^{m,n} &+& \Delta t (I - \Delta t \nabla_{U^{m}} S^{m}(U) |_{U^{m,n},U^{r,n+1}})^{-1} \\
                  &~& \left( S^{m}(U^{m,n},U^{r,n+1}) - (\nabla \cdot F^{m})^{n+1/2} - (\nabla \cdot G^{m})^{n+1/2} - (\nabla \cdot H^{m})^{n+1/2} \right) . \nonumber
\end{eqnarray} }}

\noindent
Here, $\nabla_{U^{m}} S^{m}(U)$ has the same functional form as that which was used to define the propagation operator $\mathcal{I}$ in a previous section. Therefore:
{\small {\begin{equation}
\left( I - \Delta t \nabla_{U^{m}} S^{m}(U) \right) = \left( \begin{array}{ccccc} 
1 & 0 & 0 & 0 & 0 \\
0 & 1 & 0 & 0 & 0 \\
0 & 0 & 1 & 0 & 0 \\
0 & 0 & 0 & 1 & 0 \\
\Delta t \mathbb{P} \mathbb{C} S^{E}_{\rho} & \Delta t \mathbb{P} \mathbb{C} S^{E}_{m_{x}} & \Delta t \mathbb{P} \mathbb{C} S^{E}_{m_{y}}  & \Delta t \mathbb{P} \mathbb{C} S^{E}_{m_{z}} & 1 + \Delta t \mathbb{P} \mathbb{C} S^{E}_{E} 
\end{array} \right) .
\end{equation} }}

\noindent
By inverting the above matrix, one finds:
{\small {\begin{equation}
\left( I - \Delta t \nabla_{U^{m}} S(U) \right)^{-1} = \left( \begin{array}{ccccc} 
1 & 0 & 0 & 0 & 0 \\
0 & 1 & 0 & 0 & 0 \\
0 & 0 & 1 & 0 & 0 \\
0 & 0 & 0 & 1 & 0 \\
\frac{- \Delta t \mathbb{P} \mathbb{C} S^{E}_{\rho}}{1 + \Delta t \mathbb{P} \mathbb{C} S^{E}_{E}} & \frac{- \Delta t \mathbb{P} \mathbb{C} S^{E}_{m_{x}}}{1 + \Delta t \mathbb{P} \mathbb{C} S^{E}_{E}} & \frac{- \Delta t \mathbb{P} \mathbb{C} S^{E}_{m_{y}}}{1 + \Delta t \mathbb{P} \mathbb{C} S^{E}_{E}} & \frac{- \Delta t \mathbb{P} \mathbb{C} S^{E}_{m_{z}}}{1 + \Delta t \mathbb{P} \mathbb{C} S^{E}_{E}} & \frac{1}{1 + \Delta t \mathbb{P} \mathbb{C} S^{E}_{E}} \\
\end{array} \right) .
\end{equation} }}

\noindent
The error estimate $\epsilon$ is the difference between the solution obtained from one iteration of the Picard technique where $\hat{U}$ is used as the starting value and the initial guess $\hat{U}$:
{\small {\begin{eqnarray}
\epsilon(\Delta t) = U^{m,n} &+& \frac{\Delta t}{2} \left( S^{m}(\hat{U},U^{r,n+1}) + S^{m}(U^{m,n},U^{r,n+1}) \right) \\
                             &-& \Delta t \left( ( \nabla \cdot F^{m} )^{n+1/2} + ( \nabla \cdot G^{m} )^{n+1/2} + ( \nabla \cdot H^{m} )^{n+1/2} \right) - \hat{U} . \nonumber
\end{eqnarray} }}

\noindent
Following Miniati \& Colella 2007, the correction to the initial guess is given by:
\begin{equation}
\delta(\Delta t) = \left( I - \Delta t \nabla_{U^{m}} S^{m}(U) |_{\hat{U},U^{r,n+1}} \right)^{-1} \epsilon(\Delta t) .
\end{equation}

\noindent
Therefore, the material quantities at time $t_{n+1}$ are:
\begin{equation}
U^{m,n+1} = \hat{U} + \delta(\Delta t) .
\end{equation}


\section{Conclusions and Future Work}
\noindent
This paper presents the algorithmic details for constructing a hybrid Godunov method to solve three-dimensional radiation hydrodynamical problems. Careful consideration was taken when developing this technique such that one can compute numerical solutions for a host of physical phenomena (e.g., free streaming, weak equilibrium diffusion, and strong equilibrium diffusion limits). Additionally, the algorithmic ideas in this paper were cast in such a way so that they can be easily implemented in existing codes, particularly ones that carry out MHD calculations. Future papers will showcase $(i)$ a series of multidimensional radiation hydrodynamical tests which demonstrates the robustness of the hybrid Godunov method and $(ii)$ how to combine a numerical method for radiation hydrodynamics with a technique for updating the variable tensor Eddington factor $\mathsf{f}$.


\section*{Acknowledgment}
\noindent
MDS acknowledges support from the DOE CSGF Program which is provided under grant DE-FG02-97ER25308.



\section*{Appendix 1: Partial Derivatives}
{\tiny{ \begin{eqnarray}
S^{F_{x}}_{\rho} & = & \frac{-\sigma_{t} E_{r}}{\rho^2 \mathbb{C}} \left( m_{x} + m_{x} f_{xx} + m_{y} f_{yx} + m_{z} f_{zx}  \right) - \frac{\sigma_{a} m_{x} \left( T^{4} - E_{r} \right) }{\rho^{2} \mathbb{C}} + \frac{4 \sigma_{a} m_{x} T^{3}}{\rho \mathbb{C}} \left( \gamma - 1 \right) \left( \frac{-E}{\rho^{2}} + \frac{M^{2}}{\rho^{3}} \right) , \label{eq:partfx1} \\
S^{F_{x}}_{m_{x}} & = & \frac{\sigma_{t} E_{r} \left( 1 + f_{xx} \right) }{\rho \mathbb{C}} + \frac{\sigma_{a} \left( T^{4} - E_{r} \right)}{\rho \mathbb{C}} + \frac{4 \sigma_{a} m_{x} T^{3}}{\rho \mathbb{C}} \left( \gamma - 1 \right) \left( \frac{-m_{x}}{\rho^{2}} \right) , \label{eq:partfx2} \\
S^{F_{x}}_{m_{y}} & = & \frac{\sigma_{t} E_{r} f_{yx} }{\rho \mathbb{C}} + \frac{4 \sigma_{a} m_{x} T^{3}}{\rho \mathbb{C}} \left( \gamma - 1 \right) \left( \frac{-m_{y}}{\rho^{2}} \right) , \label{eq:partfx3} \\
S^{F_{x}}_{m_{z}} & = & \frac{\sigma_{t} E_{r} f_{zx} }{\rho \mathbb{C}} + \frac{4 \sigma_{a} m_{x} T^{3}}{\rho \mathbb{C}} \left( \gamma - 1 \right) \left( \frac{-m_{z}}{\rho^{2}} \right) , \label{eq:partfx4} \\
S^{F_{x}}_{E} & = & \frac{4 \sigma_{a} m_{x} T^{3}}{\rho \mathbb{C}} \left( \gamma - 1 \right) \left( \frac{1}{\rho} \right) , \label{eq:partfx5} \\
S^{F_{y}}_{\rho} & = & \frac{-\sigma_{t} E_{r}}{\rho^2 \mathbb{C}} \left( m_{y} + m_{x} f_{xy} + m_{y} f_{yy} + m_{z} f_{zy}  \right) - \frac{\sigma_{a} m_{y} \left( T^{4} - E_{r} \right) }{\rho^{2} \mathbb{C}} + \frac{4 \sigma_{a} m_{y} T^{3}}{\rho \mathbb{C}} \left( \gamma - 1 \right) \left( \frac{-E}{\rho^{2}} + \frac{M^{2}}{\rho^{3}} \right) , \label{eq:partfy1} \\
S^{F_{y}}_{m_{x}} & = & \frac{\sigma_{t} E_{r} f_{xy} }{\rho \mathbb{C}} + \frac{4 \sigma_{a} m_{y} T^{3}}{\rho \mathbb{C}} \left( \gamma - 1 \right) \left( \frac{-m_{x}}{\rho^{2}} \right) , \label{eq:partfy2} \\
S^{F_{y}}_{m_{y}} & = & \frac{\sigma_{t} E_{r} \left( 1 + f_{yy} \right) }{\rho \mathbb{C}} + \frac{\sigma_{a} \left( T^{4} - E_{r} \right)}{\rho \mathbb{C}} + \frac{4 \sigma_{a} m_{y} T^{3}}{\rho \mathbb{C}} \left( \gamma - 1 \right) \left( \frac{-m_{x}}{\rho^{2}} \right) , \label{eq:partfy3} \\
S^{F_{y}}_{m_{z}} & = & \frac{\sigma_{t} E_{r} f_{zy} }{\rho \mathbb{C}} + \frac{4 \sigma_{a} m_{y} T^{3}}{\rho \mathbb{C}} \left( \gamma - 1 \right) \left( \frac{-m_{z}}{\rho^{2}} \right) , \label{eq:partfy4} \\
S^{F_{y}}_{E} & = & \frac{4 \sigma_{a} m_{y} T^{3}}{\rho \mathbb{C}} \left( \gamma - 1 \right) \left( \frac{1}{\rho} \right) , \label{eq:partfy5} \\
S^{F_{z}}_{\rho} & = & \frac{-\sigma_{t} E_{r}}{\rho^2 \mathbb{C}} \left( m_{z} + m_{x} f_{xz} + m_{y} f_{yz} + m_{z} f_{zz}  \right) - \frac{\sigma_{a} m_{z} \left( T^{4} - E_{r} \right) }{\rho^{2} \mathbb{C}} + \frac{4 \sigma_{a} m_{z} T^{3}}{\rho \mathbb{C}} \left( \gamma - 1 \right) \left( \frac{-E}{\rho^{2}} + \frac{M^{2}}{\rho^{3}} \right) , \label{eq:partfz1} \\
S^{F_{z}}_{m_{x}} & = & \frac{\sigma_{t} E_{r} f_{xz} }{\rho \mathbb{C}} + \frac{4 \sigma_{a} m_{z} T^{3}}{\rho \mathbb{C}} \left( \gamma - 1 \right) \left( \frac{-m_{x}}{\rho^{2}} \right) , \label{eq:partfz2} \\
S^{F_{z}}_{m_{y}} & = & \frac{\sigma_{t} E_{r} f_{yz} }{\rho \mathbb{C}} + \frac{4 \sigma_{a} m_{z} T^{3}}{\rho \mathbb{C}} \left( \gamma - 1 \right) \left( \frac{-m_{y}}{\rho^{2}} \right) , \label{eq:partfz3} \\
S^{F_{z}}_{m_{z}} & = & \frac{\sigma_{t} E_{r} \left( 1 + f_{zz} \right) }{\rho \mathbb{C}} + \frac{\sigma_{a} \left( T^{4} - E_{r} \right)}{\rho \mathbb{C}} + \frac{4 \sigma_{a} m_{z} T^{3}}{\rho \mathbb{C}} \left( \gamma - 1 \right) \left( \frac{-m_{z}}{\rho^{2}} \right) , \label{eq:partfz4} \\
S^{F_{z}}_{E} & = & \frac{4 \sigma_{a} m_{z} T^{3}}{\rho \mathbb{C}} \left( \gamma - 1 \right) \left( \frac{1}{\rho} \right) , \label{eq:partfz5} \\
S^{E}_{\rho} & = & 4 \sigma_{a} T^{3} \left( \gamma - 1 \right) \left( \frac{-E}{\rho^{2}} + \frac{M^{2}}{\rho^{3}} \right) - \frac{\left( \sigma_{a} - \sigma_{s} \right) m_{x} F_{r,x}}{\rho^{2} \mathbb{C}} + \frac{2 \left( \sigma_{a} - \sigma_{s} \right) m_{x} E_{r}}{\rho^{3} \mathbb{C}^{2}} \left( m_{x} + m_{x} f_{xx} + m_{y} f_{yx} + m_{z} f_{zx}  \right) \label{eq:parte1} \\
&~& ~~~~~~~~~~~~~~~~~~~~~~~~~~~~~~~~~~~~~~ - \frac{\left( \sigma_{a} - \sigma_{s} \right) m_{y} F_{r,y}}{\rho^{2} \mathbb{C}} + \frac{2 \left( \sigma_{a} - \sigma_{s} \right) m_{y} E_{r}}{\rho^{3} \mathbb{C}^{2}} \left( m_{y} + m_{x} f_{xy} + m_{y} f_{yy} + m_{z} f_{zy}  \right) \nonumber \\
&~& ~~~~~~~~~~~~~~~~~~~~~~~~~~~~~~~~~~~~~~ - \frac{\left( \sigma_{a} - \sigma_{s} \right) m_{z} F_{r,z}}{\rho^{2} \mathbb{C}} + \frac{2 \left( \sigma_{a} - \sigma_{s} \right) m_{z} E_{r}}{\rho^{3} \mathbb{C}^{2}} \left( m_{z} + m_{x} f_{xz} + m_{y} f_{yz} + m_{z} f_{zz}  \right)  , \nonumber \\ 
S^{E}_{m_{x}} & = & 4 \sigma_{a} T^{3} \left( \gamma - 1 \right) \left( \frac{-m_{x}}{\rho^{2}} \right) + \frac{\left( \sigma_{a} - \sigma_{s} \right) F_{r,x}}{\rho \mathbb{C}} - \frac{\left( \sigma_{a} - \sigma_{s} \right) E_{r}}{\rho^{2} \mathbb{C}^{2}} \left( 2 m_{x} + 2 m_{x} f_{xx} + m_{y} f_{yx} + m_{z} f_{zx} \right) \label{eq:parte2} \\
&~& ~~~~~~~~~~~~~~~~~~~~~~~~~~~~~~~ - \frac{\left( \sigma_{a} - \sigma_{s} \right) E_{r}}{\rho^{2} \mathbb{C}^{2}} m_{y} f_{xy} - \frac{\left( \sigma_{a} - \sigma_{s} \right) E_{r}}{\rho^{2} \mathbb{C}^{2}} m_{z} f_{xz} ,  \nonumber \\
S^{E}_{m_{y}} & = & 4 \sigma_{a} T^{3} \left( \gamma - 1 \right) \left( \frac{-m_{y}}{\rho^{2}} \right) + \frac{\left( \sigma_{a} - \sigma_{s} \right) F_{r,y}}{\rho \mathbb{C}} - \frac{\left( \sigma_{a} - \sigma_{s} \right) E_{r}}{\rho^{2} \mathbb{C}^{2}} \left( 2 m_{y} + m_{x} f_{xy} + 2 m_{y} f_{yy} + m_{z} f_{zy} \right) \label{eq:parte3} \\
&~& ~~~~~~~~~~~~~~~~~~~~~~~~~~~~~~~ - \frac{\left( \sigma_{a} - \sigma_{s} \right) E_{r}}{\rho^{2} \mathbb{C}^{2}} m_{x} f_{yx} - \frac{\left( \sigma_{a} - \sigma_{s} \right) E_{r}}{\rho^{2} \mathbb{C}^{2}} m_{z} f_{yz} , \nonumber \\
S^{E}_{m_{z}} & = & 4 \sigma_{a} T^{3} \left( \gamma - 1 \right) \left( \frac{-m_{z}}{\rho^{2}} \right) + \frac{\left( \sigma_{a} - \sigma_{s} \right) F_{r,z}}{\rho \mathbb{C}} - \frac{\left( \sigma_{a} - \sigma_{s} \right) E_{r}}{\rho^{2} \mathbb{C}^{2}} \left( 2 m_{z} + m_{x} f_{xz} + m_{y} f_{yz} + 2 m_{z} f_{zz} \right) \label{eq:parte4} \\
&~& ~~~~~~~~~~~~~~~~~~~~~~~~~~~~~~~ - \frac{\left( \sigma_{a} - \sigma_{s} \right) E_{r}}{\rho^{2} \mathbb{C}^{2}} m_{x} f_{zx} - \frac{\left( \sigma_{a} - \sigma_{s} \right) E_{r}}{\rho^{2} \mathbb{C}^{2}} m_{y} f_{zy} , \nonumber \\
S^{E}_{E} & = & 4 \sigma_{a} T^{3} \left( \gamma - 1 \right) \left( \frac{1}{\rho} \right) . \label{eq:parte5}
\end{eqnarray} }}


\section*{Appendix 2: Effective Material Eigenvectors}
{\tiny{ \begin{equation}
R^{m}_{x,\textrm{eff}} = \left( \begin{array}{ccccc} 
1                  & 1 & 0 & 0 & 1                  \\
u-a_{\textrm{eff}} & u & 0 & 0 & u+a_{\textrm{eff}} \\
v                  & v & 1 & 0 & v                  \\
w                  & w & 0 & 1 & w                  \\
\frac{V^2}{2} - u a_{\textrm{eff}} + \frac{a^{2}_{\textrm{eff}}}{\gamma-1} & \frac{V^2}{2} & v & w & \frac{V^2}{2} + u a_{\textrm{eff}} + \frac{a^{2}_{\textrm{eff}}}{\gamma-1}
\end{array} \right) ,
\end{equation}
\begin{equation}
L^{m}_{x,\textrm{eff}} = \left( \begin{array}{ccccc} 
\frac{1}{2 a_{\textrm{eff}}} \left( u + \frac{(\gamma-1) V^2}{2 a_{\textrm{eff}}} \right) & \frac{-1}{2 a_{\textrm{eff}}} \left( 1 + \frac{(\gamma-1) u}{a_{\textrm{eff}}} \right) & \frac{-1}{2 a_{\textrm{eff}}} \frac{(\gamma-1) v}{a_{\textrm{eff}}} & \frac{-1}{2 a_{\textrm{eff}}} \frac{(\gamma-1) w}{a_{\textrm{eff}}} & \frac{(\gamma-1)}{2 a^{2}_{\textrm{eff}}} \\
1 - \frac{(\gamma-1) V^2}{2 a^{2}_{\textrm{eff}}} & \frac{(\gamma-1) u}{a^{2}_{\textrm{eff}}} & \frac{(\gamma-1) v}{a^{2}_{\textrm{eff}}} & \frac{(\gamma-1) w}{a^{2}_{\textrm{eff}}} & \frac{-(\gamma-1)}{a^{2}_{\textrm{eff}}} \\
-v & 0 & 1 & 0 & 0 \\
-w & 0 & 0 & 1 & 0 \\
\frac{-1}{2 a_{\textrm{eff}}} \left( u - \frac{(\gamma-1) V^2}{2 a_{\textrm{eff}}} \right) & \frac{1}{2 a_{\textrm{eff}}} \left( 1 - \frac{(\gamma-1) u}{a_{\textrm{eff}}} \right) & \frac{-1}{2 a_{\textrm{eff}}} \frac{(\gamma-1) v}{a_{\textrm{eff}}} & \frac{-1}{2 a_{\textrm{eff}}} \frac{(\gamma-1) w}{a_{\textrm{eff}}} & \frac{(\gamma-1)}{2 a^{2}_{\textrm{eff}}}
\end{array} \right) ,
\end{equation} 
\begin{equation}
R^{m}_{y,\textrm{eff}} = \left( \begin{array}{ccccc} 
1                  & 0 & 1 & 0 & 1                  \\
u                  & 1 & u & 0 & u                  \\
v-a_{\textrm{eff}} & 0 & v & 0 & v+a_{\textrm{eff}} \\
w                  & 0 & w & 1 & w                  \\
\frac{V^2}{2} - v a_{\textrm{eff}} + \frac{a^{2}_{\textrm{eff}}}{\gamma-1} & u & \frac{V^2}{2} & w & \frac{V^2}{2} + v a_{\textrm{eff}} + \frac{a^{2}_{\textrm{eff}}}{\gamma-1}
\end{array} \right) ,
\end{equation}
\begin{equation}
L^{m}_{y,\textrm{eff}} = \left( \begin{array}{ccccc} 
\frac{1}{2 a_{\textrm{eff}}} \left( v + \frac{(\gamma-1) V^2}{2 a_{\textrm{eff}}} \right) & \frac{-1}{2 a_{\textrm{eff}}} \frac{(\gamma-1) u}{a_{\textrm{eff}}} & \frac{-1}{2 a_{\textrm{eff}}} \left( 1 + \frac{(\gamma-1) v}{a_{\textrm{eff}}} \right) & \frac{-1}{2 a_{\textrm{eff}}} \frac{(\gamma-1) w}{a_{\textrm{eff}}} & \frac{(\gamma-1)}{2 a^{2}_{\textrm{eff}}} \\
-u & 1 & 0 & 0 & 0 \\
1 - \frac{(\gamma-1) V^2}{2 a^{2}_{\textrm{eff}}} & \frac{(\gamma-1) u}{a^{2}_{\textrm{eff}}} & \frac{(\gamma-1) v}{a^{2}_{\textrm{eff}}} & \frac{(\gamma-1) w}{a^{2}_{\textrm{eff}}} & \frac{-(\gamma-1)}{a^{2}_{\textrm{eff}}} \\
-w & 0 & 0 & 1 & 0 \\
\frac{-1}{2 a_{\textrm{eff}}} \left( v - \frac{(\gamma-1) V^2}{2 a_{\textrm{eff}}} \right) & \frac{-1}{2 a_{\textrm{eff}}} \frac{(\gamma-1) u}{a_{\textrm{eff}}} & \frac{1}{2 a_{\textrm{eff}}} \left( 1 - \frac{(\gamma-1) v}{a_{\textrm{eff}}} \right) & \frac{-1}{2 a_{\textrm{eff}}} \frac{(\gamma-1) w}{a_{\textrm{eff}}} & \frac{(\gamma-1)}{2 a^{2}_{\textrm{eff}}}
\end{array} \right) ,
\end{equation} 
\begin{equation}
R^{m}_{z,\textrm{eff}} = \left( \begin{array}{ccccc} 
1                  & 0 & 0 & 1 & 1                  \\
u                  & 1 & 0 & u & u                  \\
v                  & 0 & 1 & v & v                  \\
w-a_{\textrm{eff}} & 0 & 0 & w & w+a_{\textrm{eff}} \\
\frac{V^2}{2} - w a_{\textrm{eff}} + \frac{a^{2}_{\textrm{eff}}}{\gamma-1} & u & v & \frac{V^2}{2} & \frac{V^2}{2} + w a_{\textrm{eff}} + \frac{a^{2}_{\textrm{eff}}}{\gamma-1}
\end{array} \right) ,
\end{equation}
\begin{equation}
L^{m}_{z,\textrm{eff}} = \left( \begin{array}{ccccc} 
\frac{1}{2 a_{\textrm{eff}}} \left( w + \frac{(\gamma-1) V^2}{2 a_{\textrm{eff}}} \right) & \frac{-1}{2 a_{\textrm{eff}}} \frac{(\gamma-1) u}{a_{\textrm{eff}}} & \frac{-1}{2 a_{\textrm{eff}}} \frac{(\gamma-1) v}{a_{\textrm{eff}}} & \frac{-1}{2 a_{\textrm{eff}}} \left( 1 + \frac{(\gamma-1) w}{a_{\textrm{eff}}} \right) & \frac{(\gamma-1)}{2 a^{2}_{\textrm{eff}}} \\
-u & 1 & 0 & 0 & 0 \\
-v & 0 & 1 & 0 & 0 \\
1 - \frac{(\gamma-1) V^2}{2 a^{2}_{\textrm{eff}}} & \frac{(\gamma-1) u}{a^{2}_{\textrm{eff}}} & \frac{(\gamma-1) v}{a^{2}_{\textrm{eff}}} & \frac{(\gamma-1) w}{a^{2}_{\textrm{eff}}} & \frac{-(\gamma-1)}{a^{2}_{\textrm{eff}}} \\
\frac{-1}{2 a_{\textrm{eff}}} \left( w - \frac{(\gamma-1) V^2}{2 a_{\textrm{eff}}} \right) & \frac{-1}{2 a_{\textrm{eff}}} \frac{(\gamma-1) u}{a_{\textrm{eff}}} & \frac{-1}{2 a_{\textrm{eff}}} \frac{(\gamma-1) v}{a_{\textrm{eff}}} & \frac{1}{2 a_{\textrm{eff}}} \left( 1 - \frac{(\gamma-1) w}{a_{\textrm{eff}}} \right) & \frac{(\gamma-1)}{2 a^{2}_{\textrm{eff}}}
\end{array} \right) .
\end{equation} }}


\end{document}